\pdfoutput=1

\documentclass[a4paper,11pt]{article}
\usepackage{jcappub}

\usepackage{graphicx}

\usepackage{ifthen}

\usepackage{verbatim}

\graphicspath{{figures/}}

\usepackage[normalem]{ulem}
\definecolor{DarkRed}{rgb}{0.5,0.0,0.0}
\definecolor{DarkGreen}{rgb}{0.0,0.5,0.0}
\definecolor{DarkBlue}{rgb}{0.0,0.0,0.5}
\definecolor{Magenta}{rgb}{1.0,0.0,1.0}
\definecolor{DarkMagenta}{rgb}{0.5,0.0,0.5}
\definecolor{Orange}{rgb}{1.0,0.5,0.0}
\definecolor{DarkOrange}{rgb}{0.8,0.3,0.0}
\definecolor{DarkCyan}{cmyk}{1.0,0.0,0.0,0.5}
\definecolor{Brown}{cmyk}{0.0,0.8,1,0.6}

\usepackage{soul}

\newcommand{\CS}[1]{\textcolor{DarkRed}{\textbf{[CS: #1]}}}
\newcommand{\MV}[1]{\textcolor{DarkBlue}{\textbf{[MV: #1]}}}

\usepackage{listings}

\definecolor{lstbkgdcolor}{gray}{0.85}

\lstdefinestyle{CLstyle}{language=sh,
        xleftmargin=1.5\parindent,xrightmargin=1.5\parindent,
        columns=fixed,basicstyle=\ttfamily,basewidth=0.5em,
        frame=single,
        backgroundcolor=\color{lstbkgdcolor},
        gobble=1
        }

\lstnewenvironment{snippet}{\pagebreak[2]}{}
\lstnewenvironment{CLsnippet}{\pagebreak[2]\lstset{style=CLstyle}}{\pagebreak[2]}

\newcommand{\ie}{i.e.}
\newcommand{\eg}{e.g.}

\newcommand{\refeqn}[2][eqn:]{eq.~(\ref{#1#2})}
\newcommand{\Refeqn}[2][eqn:]{Equation~(\ref{#1#2})}
\newcommand{\reftab}[2][tab:]{table~\ref{#1#2}}

\newcommand{\reffig}[2][fig:]{figure~\ref{#1#2}}
\newcommand{\Reffig}[2][fig:]{Figure~\ref{#1#2}}
\newcommand{\refsec}[2][sec:]{section~\ref{#1#2}} 
\newcommand{\refapp}[2][sec:]{the appendix}
\newcommand{\Refapp}[2][sec:]{The appendix}

\newcommand{\ifmulticol}[2]{%
  \ifthenelse{\lengthtest{1.9\columnwidth<\textwidth}}{#1}{#2}%
}

\newcommand{\insertwidefig}[2][\widefigwidth]{%
    \hspace*{\stretch{1}}
    \includegraphics[keepaspectratio,width=#1\textwidth,
                     height=0.75\textheight]{#2}
    \hspace*{\stretch{1}}
}

\newcommand{\erf}{\mathop{\mathrm{erf}}}

\newcommand{\lae}{%
  \ensuremath{\raisebox{-0.4ex}{%
    $\,\buildrel{\scriptstyle<}\over{\scriptstyle\sim}\,$}%
    }%
  }

\providecommand{\lesssim}{\lae}

\newcommand{\midtilde}{{\raise.17ex\hbox{$\scriptstyle\sim$}}}

\newcommand{\scinot}[2]{\ensuremath{#1\times\!\!10^{#2}}}

\newcommand{\mchi}{\ensuremath{m}}
\newcommand{\rhochi}{\ensuremath{\rho_{0}}}
\newcommand{\nchi}{\ensuremath{n_{0}}}

\newcommand{\bv}{\ensuremath{\mathbf{v}}}

\newcommand{\bV}{\ensuremath{\mathbf{V}}}

\newcommand{\vmin}{\ensuremath{v_\mathrm{min}}}
\newcommand{\vmp}{\ensuremath{v_0}}
\newcommand{\vrot}{\ensuremath{v_\mathrm{rot}}}

\newcommand{\bvobs}{\ensuremath{\mathbf{v}_\mathrm{obs}}}
\newcommand{\bvbulk}{\ensuremath{\mathbf{v}_\mathrm{MB}}}
\newcommand{\vesc}{\ensuremath{v_\mathrm{esc}}}
\newcommand{\Nesc}{\ensuremath{N_\mathrm{esc}}}

\newcommand{\eone}{\ensuremath{\hat{\boldsymbol{\varepsilon}}_1}}  
\newcommand{\etwo}{\ensuremath{\hat{\boldsymbol{\varepsilon}}_2}}  

\newcommand{\MB}{\ensuremath{\mathrm{MB}}}      

\newcommand{\fgal}{\ensuremath{\widetilde{f}}}        
\newcommand{\fMB}{\ensuremath{f_{\MB}}}               
\newcommand{\fearth}{\ensuremath{f_{\oplus}}}         

\newcommand{\qmax}{\ensuremath{q_{\mathrm{max}}}}  

\newcommand{\fpSI}{\ensuremath{f_{\mathrm{p}}}}
\newcommand{\fnSI}{\ensuremath{f_{\mathrm{n}}}}

\newcommand{\sigmaSI}{\ensuremath{\sigma_{0}}}
\newcommand{\sigmapSI}{\ensuremath{\sigma_{\mathrm{p}}}}

\newcommand{\mup}{\ensuremath{\mu_{\mathrm{p}}}}

\newcommand{\msun}{\ensuremath{\mathrm{M}_{\odot}}}

\newcommand{\kms}{\ensuremath{\mathrm{km\,s^{-1}}}}

\newcommand{\DMHaloCalc}{\texttt{DMHaloCalc}}

\begin{document}


\title{The impact of baryons on the direct detection of dark matter}


\author[a]{Chris Kelso}
\author[b]{Christopher Savage}
\author[c]{Monica Valluri}
\author[b,d]{Katherine Freese}
\author[e]{Gregory S.\ Stinson}
\author[f,g]{Jeremy  Bailin}

\affiliation[a]{
  Department of Physics,
  University of North Florida,
  Jacksonville, FL 32224, USA}
\affiliation[b]{
  Nordita,
  KTH Royal Institute of Technology and Stockholm University,
  SE-106 91 Stockholm, Sweden}
\affiliation[c]{
  Department of Astronomy,
  University of Michigan,
  Ann Arbor, MI 48109, USA}
\affiliation[d]{
  Department of Physics,
  University of Michigan,
  Ann Arbor, MI 48109, USA}
\affiliation[e]{
  Max-Planck-Institut f\"ur Astronomie,
  K\"onigstuhl 17,
  D-69117, Heidelberg, Germany}
\affiliation[f]{
  Department of Physics \& Astronomy, 
  University of Alabama, 
  Tuscaloosa, AL 35487, USA}
\affiliation[g]{
  National Radio Astronomy Observatory,
  P.O. Box 2,
  Green Bank, WV 24944, USA}
\emailAdd{chris@savage.name}
\emailAdd{ckelso@unf.edu}
\emailAdd{mvalluri@umich.edu}
\emailAdd{ktfreese@umich.edu}
\emailAdd{stinson@mpia.de}
\emailAdd{jbailin@ua.edu}
\abstract{ 
The spatial and velocity distributions of dark matter particles in the Milky Way
Halo affect the signals expected to be observed in searches for dark matter.
Results from direct detection experiments are often analyzed assuming a
simple isothermal distribution of dark matter, the Standard Halo Model
(SHM). Yet there has been skepticism regarding the validity of this simple model due to
the complicated gravitational collapse and merger history of actual galaxies.  In this paper we compare the SHM to the results of cosmological hydrodynamical simulations of galaxy
formation to investigate whether or not  the SHM is a good representation of the true WIMP distribution in the analysis of direct detection data.  
We examine two  Milky Way-like galaxies from the MaGICC cosmological simulations  (a) with dark matter only and (b) with baryonic physics included. The inclusion of baryons drives the shape of the DM halo to become more spherical and makes the velocity distribution of dark matter particles less anisotropic especially at large heliocentric velocities, thereby making the SHM a better fit.  We also note that we do not find a significant disk-like rotating dark matter component in either of the two galaxy halos with baryons that we examine, suggesting that dark disks are not a generic prediction of cosmological hydrodynamical simulations.  We conclude that in the Solar neighborhood, the SHM is in fact a good approximation to the true dark matter distribution in these cosmological simulations (with baryons) which are reasonable representations of the Milky Way, and hence can also be used for the purpose of dark matter direct detection calculations.

%
} 



\maketitle
\flushbottom





\section{Introduction}
\label{sec:Intro}

The Milky Way, along with other galaxies, is believed to be
encompassed in a massive dark matter halo of unknown composition.  
Only 5\% of the Universe consists of normal matter (baryons), while the
remainder is  dark matter (23\%) and dark energy (72\%) \cite{Komatsu:2010fb,Ade:2015xua}.  Identifying the
nature of this dark matter is one of the longest outstanding problems
in modern physics.  Leading candidates for this dark matter are
Weakly Interacting Massive Particles (WIMPs), a generic class of
particles that includes the lightest supersymmetric particle.
These particles undergo weak
interactions and their expected masses range from 1~GeV to 10~TeV.
If present in thermal equilibrium in the early
universe, these particles annihilate with one another so that a predictable number of
them remain today.  For a wide range of parameters, the relic density of these particles is found to be
roughly in agreement with
the value measured by the Wilkinson Microwave Anisotropy Probe (WMAP) and Planck satellite.

Thirty years ago, Refs. \cite{Drukier:1983gj, Goodman:1984dc} proposed
that a laboratory mechanism for detecting weakly interacting particles, 
including WIMPs, via coherent scattering with nuclei.  Soon after,  the dark matter 
detection rates in the context of a Galactic Halo of WIMPs were computed for the first time, and
it was proposed that they could be differentiated from background by looking for annual modulation of the signal~\cite{Drukier:1986tm}. 
 Since that time, a multitude of experimental efforts to detect WIMPs has been underway, with one currently claiming detection.  The basic goal of direct detection experiments is to measure the energy deposited when weakly interacting particles  scatter off of nuclei in the detector, depositing $\sim$keV in the nucleus.

From the beginning of dark matter phenomenology, the fiducial model assumed for the dark matter distribution in the Galaxy
 has been the Standard Halo Model (SHM) \cite{Drukier:1986tm}. 
  The SHM is an isothermal spherical dark matter distribution \cite{Drukier:1986tm},
which leads to a Maxwell-Boltzmann distribution with dispersion velocity equal to
 the (circular) disk rotation speed in the solar neighborhood (see Eq.(\ref{eqn:TMB}) and the discussion below it)\footnote{In the SHM the
 Maxwellian distribution is then truncated at the escape velocity from the Milky Way, since WIMPs with sufficiently high velocity 
escape the Galaxy's potential well and the high-velocity tail of the distribution is depleted.  Most studies of dark matter direct
detection have assumed this SHM.}.

Yet, in the community there has been the belief that this simple SHM cannot be trusted as it does not adequately capture the 
true velocity distribution of the Galactic WIMPs. Dark matter halos in cosmological simulations (with dark matter
only)
were known to be triaxial and have anisotropic velocity distributions owing to their complex merger history. The anisotropic velocity distributions have been shown to deviate significantly from the SHM~\cite{Green:2011bv} and alternative analytic fits that were thought to better match the true distribution, particularly at the high-velocity tail have been explored~\cite{Lisanti:2010qx,Mao:2012hf}.  These alternatives to the SHM were motivated in part 
by results of dark matter only N-body simulations of structure formation which are limited by the fact that they do not include baryonic physics.  In this work we examine the results of simulations that do include baryons.   
It is the purpose of this paper to investigate whether or not 
the SHM is a good representation of the true WIMP distribution in these simulations for the analysis of direct detection data.  
 
The process of galaxy formation proceeds hierarchically via the  clustering and aggregation of dark matter halos. After the first dark matter halos form, gas begins to be gravitationally attracted into them where it cools, condenses and forms the first generations of stars.   As a galaxy gains mass (by accreting smaller galaxies) the dark matter halos also grow.  Consequently galaxies like the Milky Way are embedded in dark matter halos that formed via the repeated merger of protogalactic fragments consisting of dark matter, gas, and early generations of stars. 

The measurement of the shapes of dark matter halos is challenging and the shape of the Milky Way Halo is currently uncertain \citep[e.g.][]{Law:2010pc,Banerjee:2011rr,Loebman:2014xha}.  
Since dark matter constitutes about 85\% of the total mass in the Universe, most early simulations of galaxy formation relied purely on collisionless simulations of dark matter particles. These early collisionless $N$-body simulations showed that hierarchical
growth results in dark matter halos whose  shapes are highly triaxial or prolate \citep{Dubinski:1991bm,
Jing:2002np,Bailin:2004wu,Allgood:2005eu,Butsky:2016a} with radially anisotropic velocity distributions \citep{Colin:1999bh,Hansen:2005yj}.

More recently, N-body simulations have been performed that take into consideration not only collisionless
dark matter particles but also the effects of the condensation of baryons.  These simulations find somewhat different
halo shapes 
that are more spherical or axisymmetric \citep[]{Dubinski:1991bm,Kazantzidis:2004vu,Debattista:2007yz,Tissera:2009cm,
Kazantzidis:2010jp}. When baryons are taken into account, analysis of the orbital properties of dark matter particles in both idealized simulations \citep{Valluri:2009ir,Valluri:2011sn} and cosmological hydrodynamical simulations \citep{Valluri:2013tj} show that the orbits  remain box-like or chaotic (characteristic of triaxial halos) at the very center of the halo even though the halo is nearly spherical, but at intermediate radii (similar to the location of the Sun in the Milky Way) the orbits have more angular momentum and are comprised of tube-like orbits. 
  
There has been much attention paid to the possibility of a `dark disk' in the Milky Way either produced as a result of preferential accretion of satellites in the plane of the baryonic disk which results in an enhanced density of dark matter in this region \citep{Read:2008,Read:2008fh}, or due to non-spherical adiabatic contraction of a more spherical dark matter halo in response to the growth of a baryonic disk \citep{Piffl:2015xua}.  A recent study of the metallicity distribution of stars in the Milky Way disk from the Gaia-ESO survey \citep{Ruchti:2015bja} finds no observational evidence for the populations of stars that are expected to be brought into the disk by satellites that would be required to create a dark disk, suggesting the Milky Way halo had few in-plane accretion events. 
 
A previous study of a disk galaxy from a cosmological hydrodynamical simulation found that the dark matter in the solar neighborhood had a significant amount of rotation similar to a  `dark disk' \citep{Ling:2009eh}.  We do not find a significant disk-like rotating dark matter component in either of the two galaxy halos with baryons that we examine, suggesting that dark disks are not a generic prediction of cosmological hydrodynamical simulations.
 
 In this paper we study a new generation of high resolution cosmological hydrodynamical and N-body simulations designed to mimic the formation of disk galaxies similar in size and global properties to the Milky Way. This state-of-the-art suite of simulations includes baryonic physics that results in  disk galaxies which have flat rotation curves, exponential surface brightness profiles, and match a wide range of observed disk scaling relationships \citep{Brook:2012fm}.  We study two different galaxies from the MaGICC simulations \citep{Stinson:2012uh}. For one of these simulations we also examine a dark-matter only simulation that was generated from the same set of dark matter initial conditions and with the same cosmology to enable us to quantify the effects of baryons on the dark matter detection signal. The simulations are described in detail in Section~\ref{sec:Simulations}.
 
  While the orbits of dark matter particles in simulations of dark matter halos with baryonic condensation are different from orbits in idealized singular isothermal spheres, it is unclear if these differences are significant enough in the solar neighborhood that they would  impact dark matter direct detection experiments.  It is the goal of this paper to address this question.



\section{Simulations}
\label{sec:Simulations}

The cosmological hydrodynamical simulations we use in this paper are from the Making Galaxies in a Cosmological Context \citep[MaGICC,][]{Stinson:2012uh}. The initial conditions assume a Lambda Cold Dark Matter ($\Lambda$CDM), WMAP3 cosmology with $H_0 = 73$ \kms Mpc$^{-1}$, $\Omega_m$ = 0.24, $\Omega_\Lambda$ = 0.76, $\Omega_b$ = 0.04 and $\sigma_8 = 0.79$ \citep{Spergel:2006hy}. A sample of 16 galaxy halos with masses between $\sim 5 \times 10^{11}$  to $\sim 2\times 10^{12}$\msun\, at redshift zero were chosen at random from a dark-matter only simulation \citep{Stinson:2010xe}. The selection criteria required that for each selected halo (at $z=0$) there was no structure within 2.7 Mpc with a mass greater than $\sim 5 \times 10^{11}$\msun.  Once the galaxies were chosen from the DM only runs,
 the simulations for those regions containing the galaxies were re-evolved from high redshift with high resolution, this time with baryonic physics included.  The simulation volume was large enough to ensure a realistic angular momentum distribution and merger history. In order to achieve sufficient mass and spatial resolution, the simulations employ a common zooming technique that adds high resolution particles in the region of interest while following other regions with much lower resolution particles. In the highest resolution region of each simulation the dark matter, gas and star particles have masses of 1.11$\times 10^6$\msun, 2.2$\times 10^5$\msun\, and $< 6.3\times 10^4$\msun\, respectively, and a gravitational softening length of 310 pc. The simulation was carried out using the SPH code GASOLINE \citep{Wadsley:2003vm} and includes low-temperature metal cooling \citep{Shen:2009zd} a Schmidt-Kennicutt star formation law \citep{Kennicutt:1997ng} and UV background radiation. A prior generation of simulations to MaGICC called McMaster Unbiased Galaxy Simulations  \citep[MUGS,][]{Stinson:2010xe} uses the same initial conditions and cosmology but has a different implementation of the stellar feedback. Both MUGS and MaGICC employ the `blast wave' model of supernova feedback, where gas cooling is locally suspended in order to mimic the thermal heating of gas from supernovae \citep{Stinson:2006cp}. MaGICC  differs from MUGS in that it also includes early energy input into the ISM, from massive stars. This early feedback heats the surrounding gas from the moment a star forms, rather than waiting for the first core collapse supernovae which are triggered $\sim 4$ Myr after the star formed \citep{Stinson:2012uh}. As a consequence of the early feedback implemented in MaGICC, the simulated galaxies we use in this paper suffer much less from overcooling and have small (more realistic) bulges. 

In this paper we focus on the dark matter halos of two disk galaxies from the MaGICC simulations with global properties  similar to that of the Milky Way. These are called `g1536' and `g15784'. The latter has a more massive bulge and a larger total mass and consequently a more rapidly rising rotation curve.  g15784 has a more active accretion history with its last major merger at $z=2$ while g1536 has its most recent merger at $z=2.9$. The use of two disk galaxies both broadly resembling the Milky Way, evolved with the same baryonic physics, but  arising from different initial conditions and having different accretion histories, allows us to assess how sensitive our results are to the details of  the galaxy assembly process. In addition to these two fully hydrodynamical simulations of disk galaxies, we also analyze a dark matter only simulation  which was generated with cosmological initial conditions identical to g1536, which we refer to  as `g1536DM'. The baryonic physics (especially early feedback which unbinds a number of the lower mass subhalos and baryonic compression) alters the virial mass, virial radius, shape of the dark matter halo and its phase space distribution function. A direct comparison of the dark matter detection rates in `g1536' and `g1536DM' enables us to directly assess how these rates might be affected in the Milky Way, since both simulations arise from the same cosmological initial conditions. 

Since all three simulations have slightly different total matter distributions, their rotation curves differ from each other and from the rotation curve of the Milky Way. In  \reftab{Simulations} we provide additional details on the three different simulations and the properties of the spatial and velocity distributions of DM particles therein. 
\begin{table}
  \begin{center}
  \begin{tabular}{llccc}
    \hline \hline 
      & & \textbf{g1536DM} & \textbf{g1536} & \textbf{g15784} \\
      & & (DM-only)        & (DM+baryons)   & (DM+baryons)    \\
    \hline 
    \multicolumn{2}{l}{Simulation}
      & & & \\
   & Virial Mass [M$_{\odot}$]              & \scinot{7.48}{11}  & \scinot{5.84}{11}  & \scinot{1.50}{12}  \\
   & Virial Radius [kpc]                          &           260          &            143           &              328        \\
    & DM particle mass [M$_{\odot}$]        & \scinot{1.33}{6}  & \scinot{1.11}{6}  & \scinot{1.11}{6}  \\
    & Circular velocity (at $R=8$~kpc) \makebox[0pt][l]{[km/s]}
                                            & 108               & 187               & 273               \\
    \multicolumn{2}{l}{Torus ($r_1=8$~kpc, $r_2=2$~kpc)} & & & \\
    & Number of DM particles                          & 3085              & 4849              & 6541              \\
    & Average DM density [GeV/cm$^3$]       & 0.270             & 0.346             & 0.493             \\
    & Average velocity ($U$,$V$,$W$) [km/s] & $(0.0,5.3,-0.5)$  & $(2.7,21.6,2.3)$  & $(0.9,18.5,3.4)$  \\
    & Velocity s.d.\ ($\sigma_U$,$\sigma_V$,$\sigma_W$) [km/s]   & $(109,95,90)$    & $(144,128,121)$   & $(205,166,177)$   \\
    & RMS speed [km/s]                      & $98\;\!\sqrt{3}$  & $133\;\!\sqrt{3}$ & $184\;\!\sqrt{3}$ \\
    & Maximum speed [km/s]                  & 359               & 454               & 600.              \\
    \hline \hline 
  \end{tabular}
  \end{center}
  \caption[Simulations]{%
    Properties of largest dark matter halos in the three $N$-body simulation used in this work. We use the definition of the virial radius of Ref~\cite{Bryan:1997dn} which corresponds approximately to the radius within which the average density of the halo is $\sim$100 times the critical density.
    }
  \label{tab:Simulations}
\end{table}

The DM halos with baryons have significantly different shapes from the halo in the dark matter-only simulation.  Halo shapes are measured in terms of the axis ratios of the mass distribution stratified on concentric ellipsoids. The axis ratio $b/a$ is the ratio of the intermediate to long axis scale lengths and $c/a$ is the ratio of the short to long axis scale lengths. The shape of the dark matter distribution in ellipsoidal shells is computed using
shells centered on the minimum of the potential. The shape is determined at each
radius $r$ by finding the ellipsoidal shell of width $\sim 1.2$~kpc
whose principal axes $\mathbf{a}$, $\mathbf{b}$, and $\mathbf{c}$
self-consistently (1) have a geometric mean radius $\sqrt{abc}=r$, and
(2) are the eigenvectors of the second moment tensor of the mass
distribution within the ellipsoidal shell. In practice, this is done by starting with a spherical shell, calculating the second moment tensor, deforming the shape of
the shell to match the eigenvectors of the tensor, and iterating until
the solution has converged; this is very similar to the method
advocated by \citep{Zemp:2011ed}. 

\begin{figure*}
  \insertwidefig{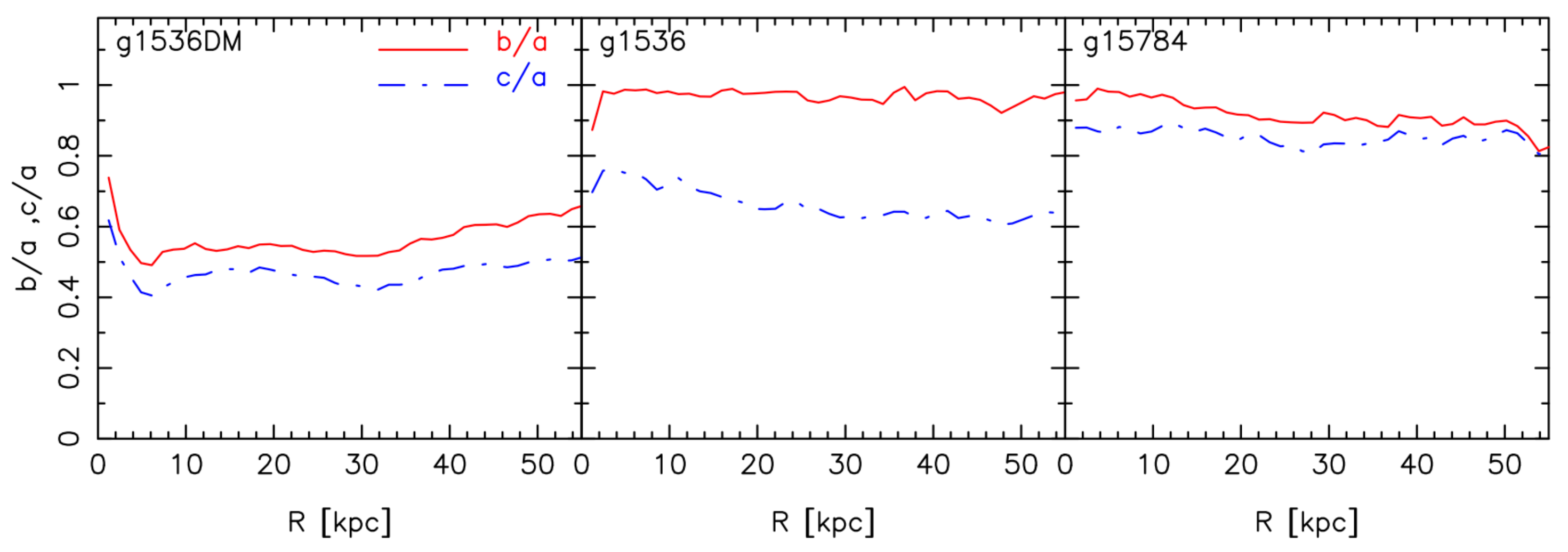}
  \caption{
    \footnotesize
Shapes of triaxial dark matter halos as a function of the ellipsoidal radius $R=\sqrt{(x/a)^2+(y/b)^2+(z/c)^2}$  as parametrized by short axis scale length/long axis scale length ($c/a$) and intermediate axis scale length/long axis scale length ($b/a$) for the three halos studied in this work.  In g1536DM the halo is highly prolate with $c/a \sim 0.45$ and $b/a \sim 0.55$.  With baryons added (middle panel), this halo becomes close to axisymmetric $b/a\sim1$, but remains somewhat flattened ($b/c \sim 0.75$). g15784 (right panel, baryonic physics included) is nearly spherical within 50~kpc. 
    }
  \label{fig:haloshape}
\end{figure*}

\Reffig{haloshape} shows the halo shape axis ratios as a function of the ellipsoidal radius $R=\sqrt{(x/a)^2+(y/b)^2+(z/c)^2}$ for the three different halos studied in this work.  In the DM only simulation (\reffig{haloshape}~left) $c/a \sim 0.45$ and $b/a \sim 0.55$ indicating that this halo is highly prolate (football shaped) except in the inner most region ($\sim$2~kpc). In contrast, g1536 with baryons is close to axisymmetric $b/a\sim1$, but quite flattened (0.6<$b/c<0.8$), while g15784 has  $0.9<b/a<1$ and $c/a\sim0.9$ indicating that it is is nearly spherical within 50~kpc. g1536DM is the most triaxial because of the absence of baryons \citep[]{Dubinski:1991bm,Kazantzidis:2004vu,Debattista:2007yz,Tissera:2009cm,
Kazantzidis:2010jp}, while the differences in the shapes of g1536 and g15784 reflects the differences in their cosmological accretion histories. Comparison of these three halos therefore enables us to assess the general applicability of our results.  It is also interesting to note that in all cases, both $c/a$ and $c/b$ are relatively constant as a  function of radius (which is not true of all cosmological simulations \citep{Zemp:2011ed}, but is not unique to these simulations \citep{Loebman:2014xha}).  

\section{Local Dark Matter Distribution}
\label{sec:Local}

Of relevance to direct detection experiments is the amount and velocity
distribution of the dark matter particles in the neighborhood of the
Earth, with the overall rate of WIMP-nucleus interactions proportional
to the former and the energy spectrum of the interactions dependent on
the latter (faster WIMPs can transfer more energy in a collision with
a nucleus).  Though the N-body simulations are not precisely
representative of the Milky Way, they are qualitatively very similar and
thus useful conclusions can be drawn from examining direct detection
signals expected in such halos.  For each simulation we compute the detection
signals in the frame-of-reference of a planet with Earth-like motion relative to the Galactic center: orbiting a star at 30~km/s with the orbital plane
inclined  at 60$^\circ$ to the Galactic disk plane, and  with the star located in and moving with the stellar rotation velocity at a distance of 8~kpc from the Galactic center\footnote{The displacement of the Sun from the Galactic plane is neglected because the particle smoothing length in the simulation is a factor of 10 larger than this.}.
For the g1536DM simulation, which contains no galactic disk, the disk
plane is assumed to be that for which the net angular momentum vector
is normal, which is close to where a disk would be expected to form \citep{Aumer:2012uf}.

Due to the approximate axial symmetry of the galaxies, the distribution of dark matter in the local neighborhood can be characterized by the distribution of simulation particles in all spheres centered on possible locations of Earth within the Galaxy. This corresponds to a torus with a major radius of 8~kpc (the Solar circle); we take the minor radius to be 2~kpc.
  The tori in the two simulations that contain both dark matter and baryons, g1536 and g15784, contain 4849 and 6541 particles, respectively, while the torus in the dark matter only simulation (g1536DM) contains 3085 particles.  The mean velocities and standard deviations  of the particles within these tori are shown in \reftab{Simulations}.

\begin{figure*}
  \includegraphics[trim=50.pt 110.pt 100.pt 160.pt,clip,width=0.31\textwidth]{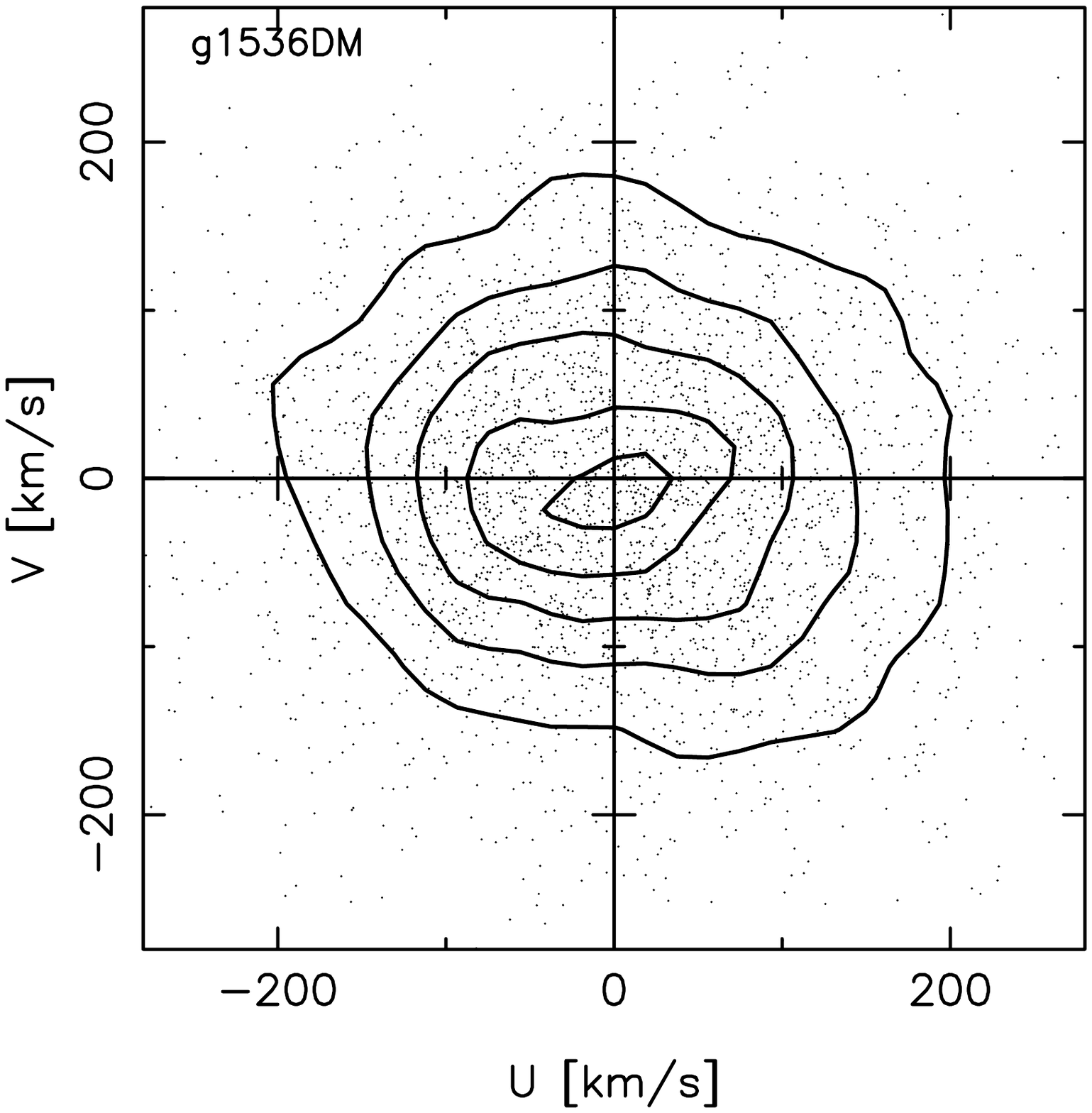}
   \includegraphics[trim=50.pt 110.pt 100.pt 160.pt,clip,width=0.31\textwidth]{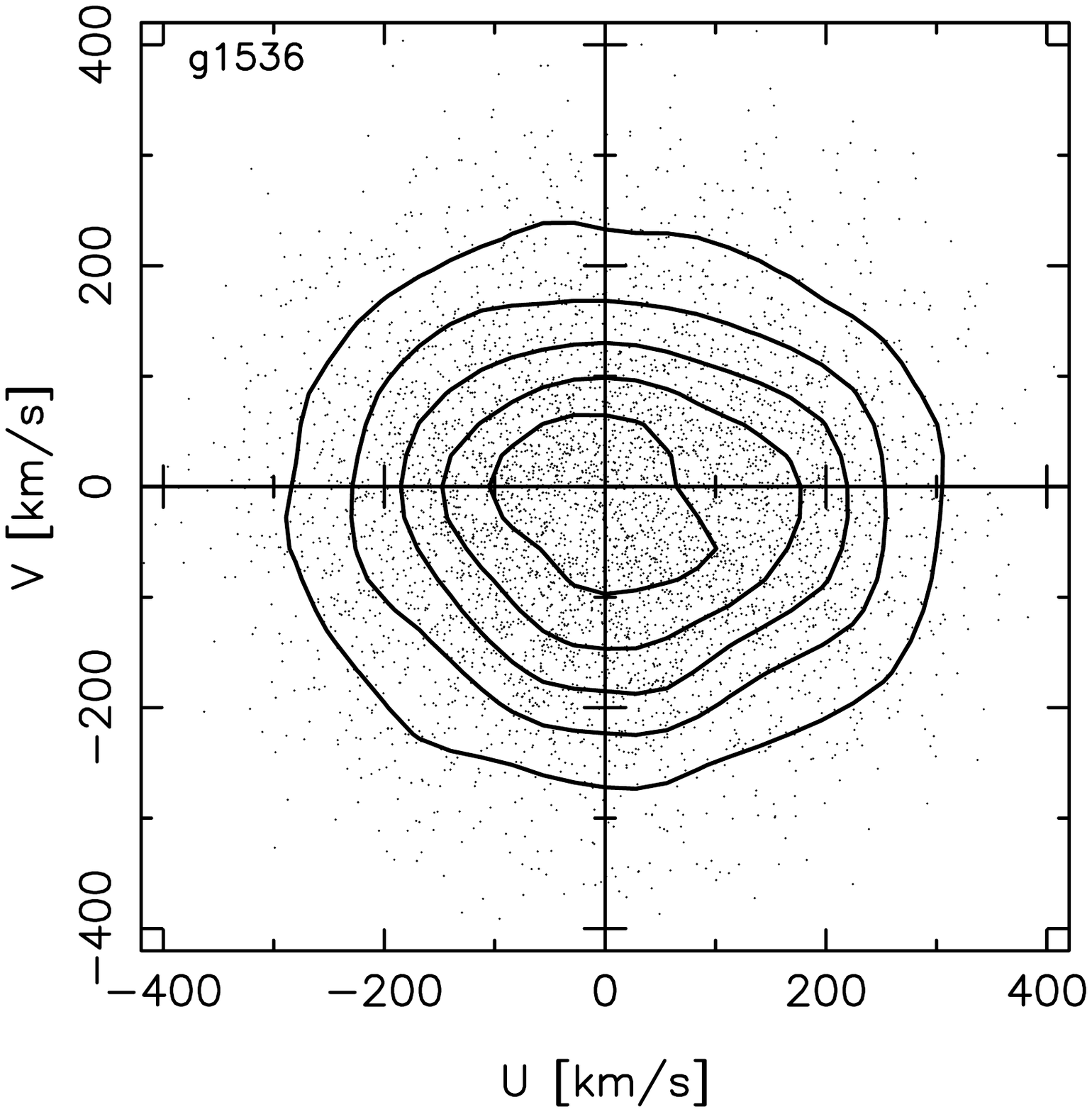}
    \includegraphics[trim=50.pt 110.pt 100.pt 160.pt,clip,width=0.31\textwidth]{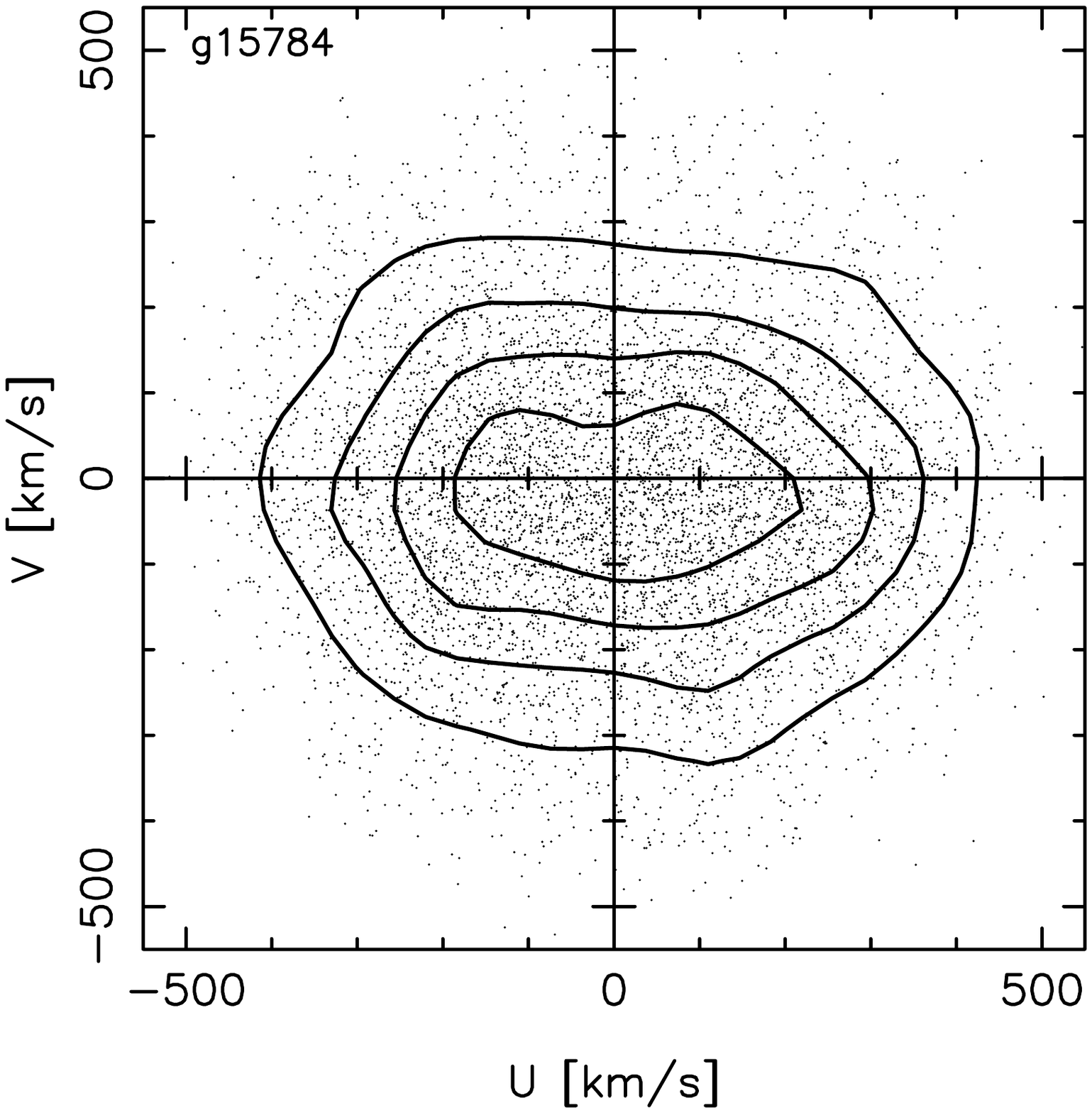}
    \includegraphics[trim=50.pt 110.pt 100.pt 160.pt,clip,width=0.31\textwidth]{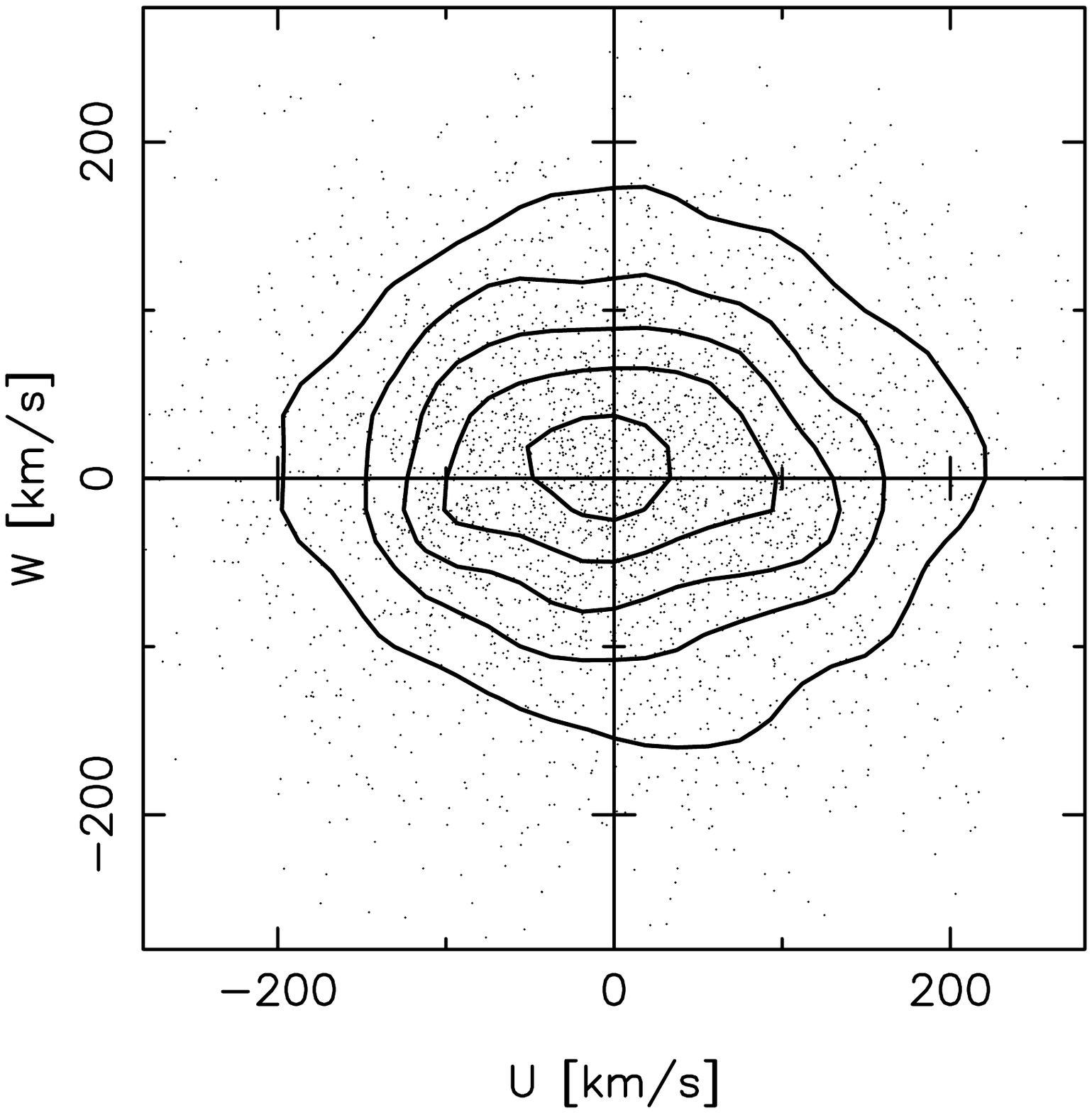}
   \includegraphics[trim=50.pt 110.pt 100.pt 160.pt,clip,width=0.31\textwidth]{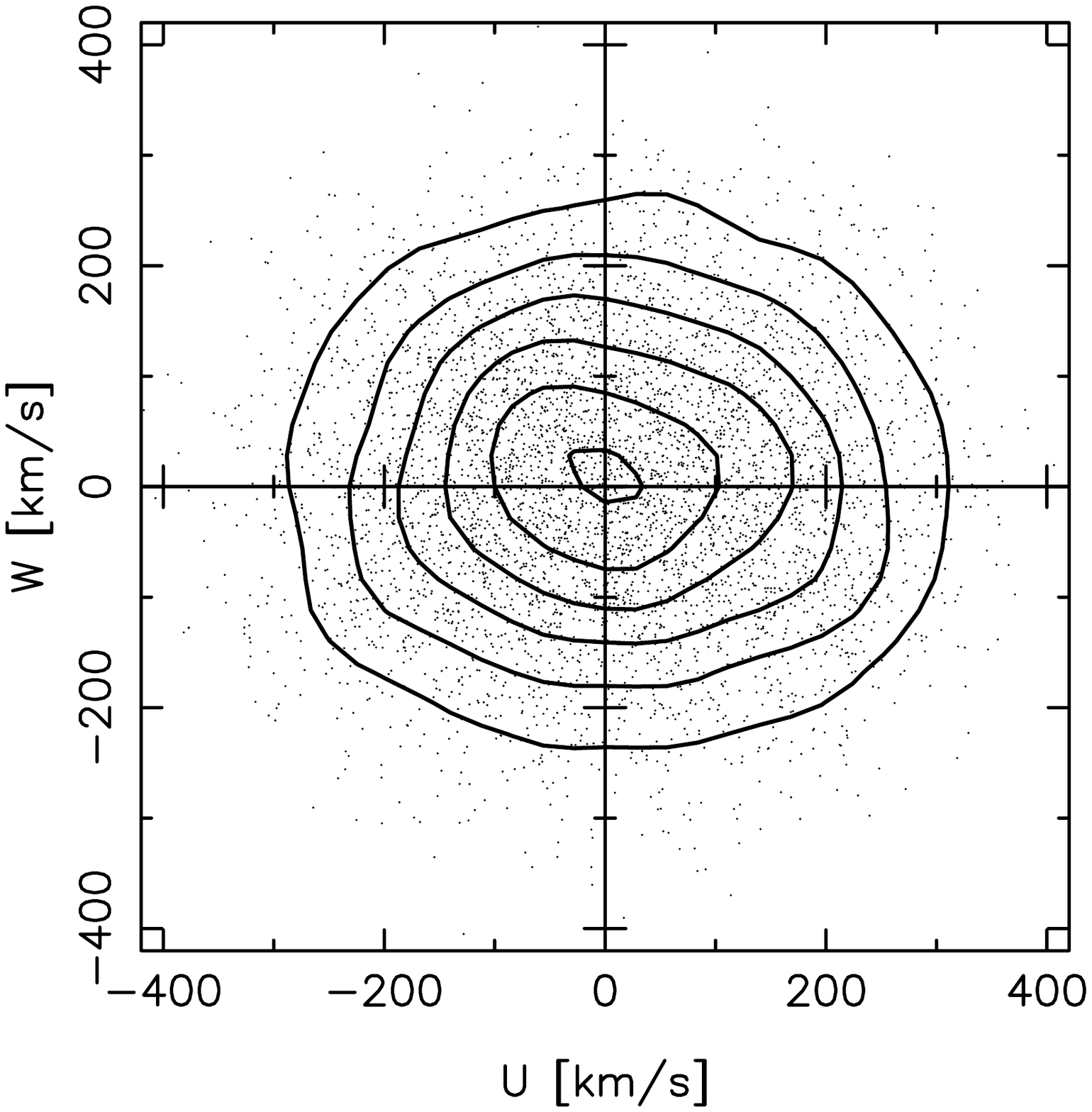}
    \includegraphics[trim=50.pt 110.pt 100.pt 160.pt,clip,width=0.31\textwidth]{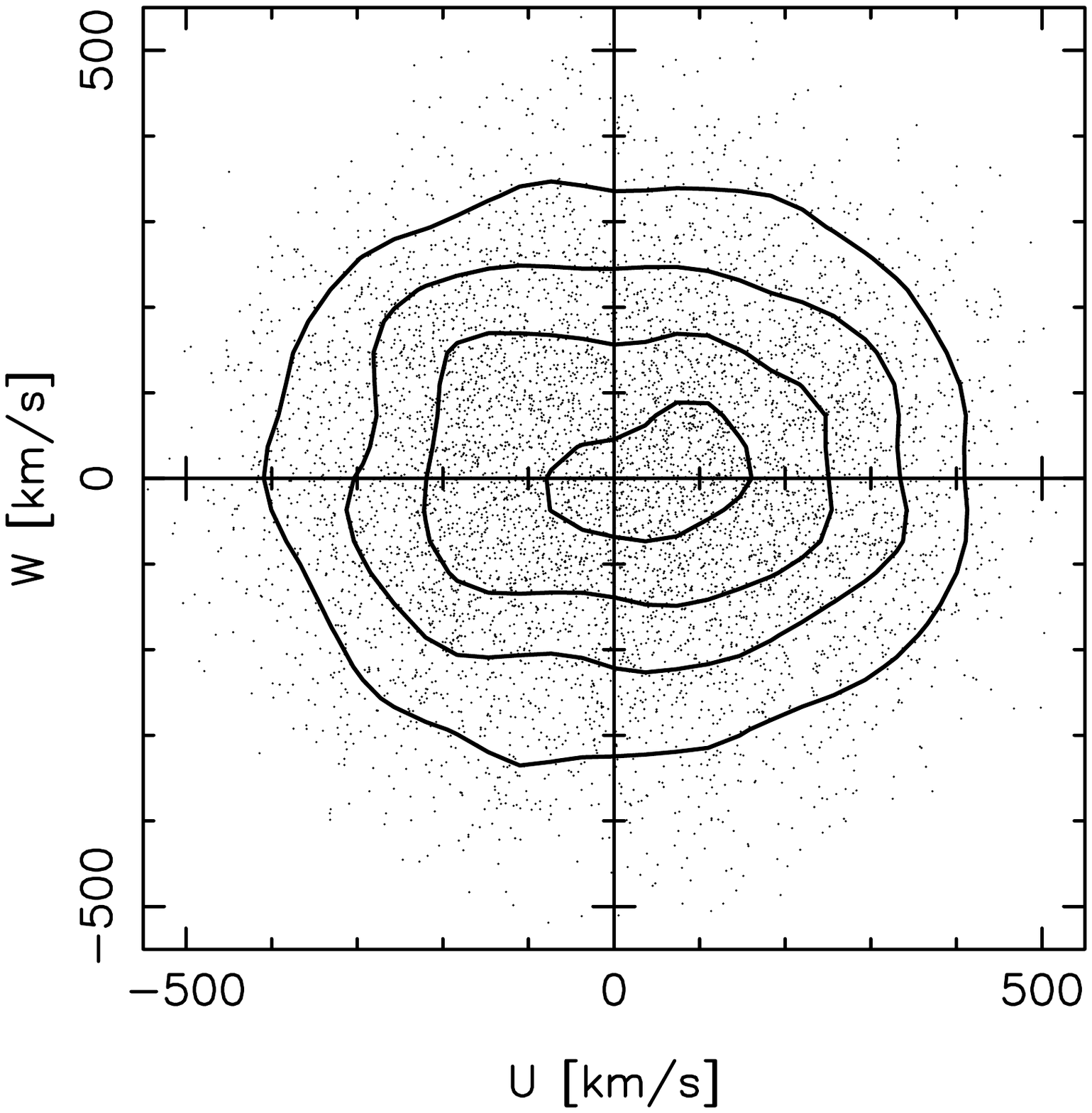}
    \includegraphics[trim=50.pt 110.pt 100.pt 160.pt,clip,width=0.31\textwidth]{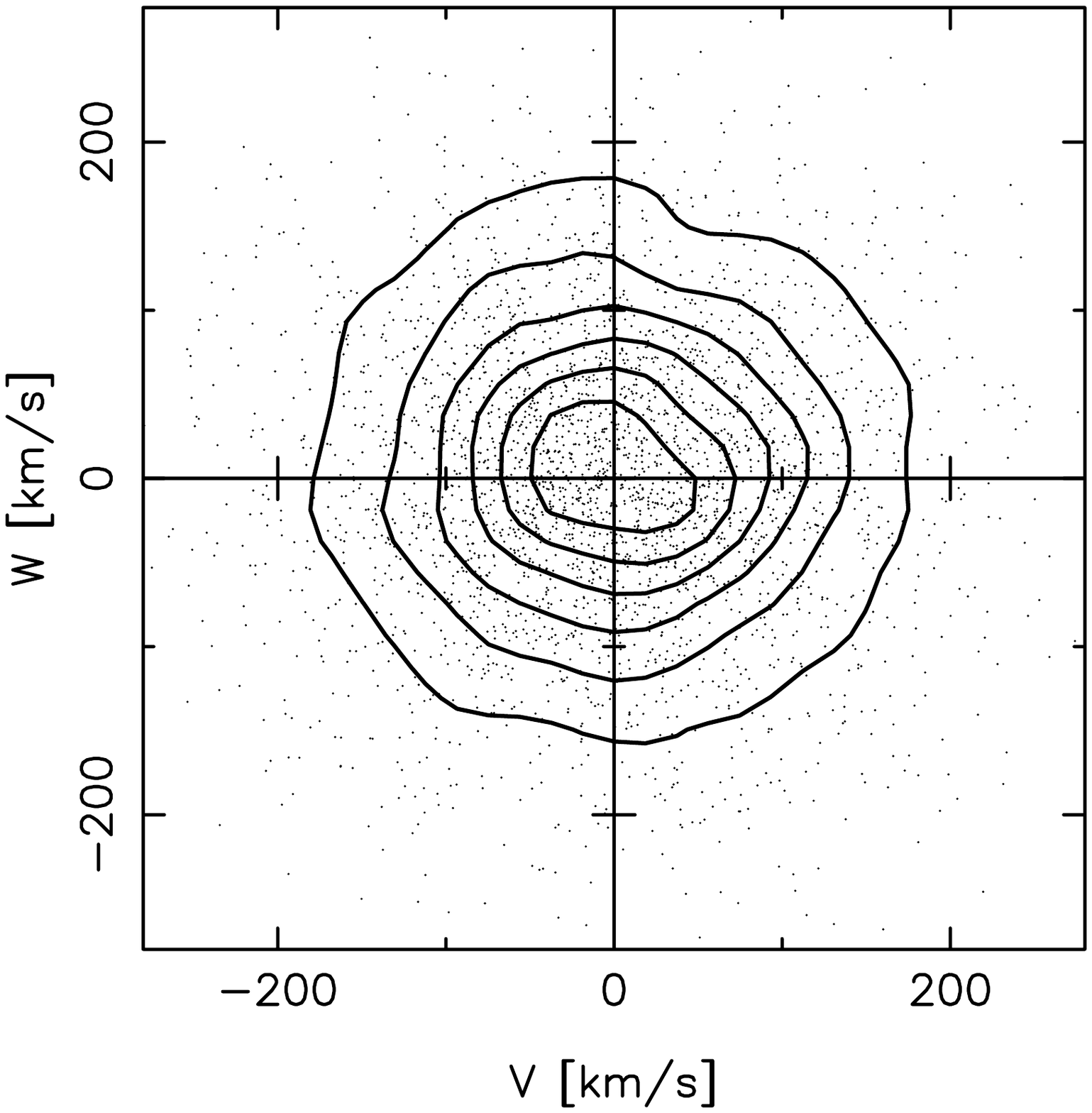}\hspace{0.37cm}
   \includegraphics[trim=50.pt 110.pt 100.pt 160.pt,clip,width=0.31\textwidth]{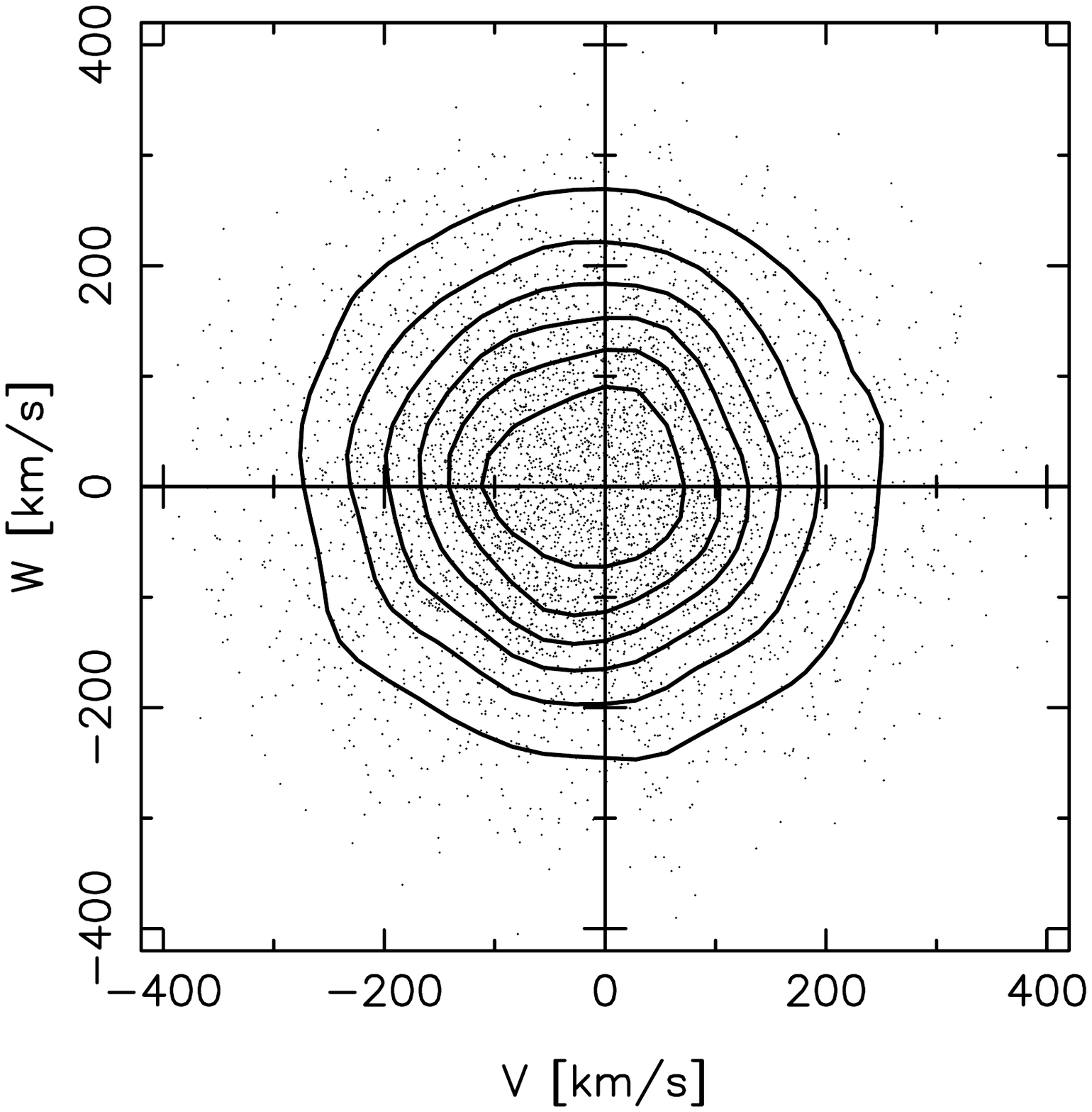}\hspace{0.37cm}
    \includegraphics[trim=50.pt 110.pt 100.pt 160.pt,clip,width=0.31\textwidth]{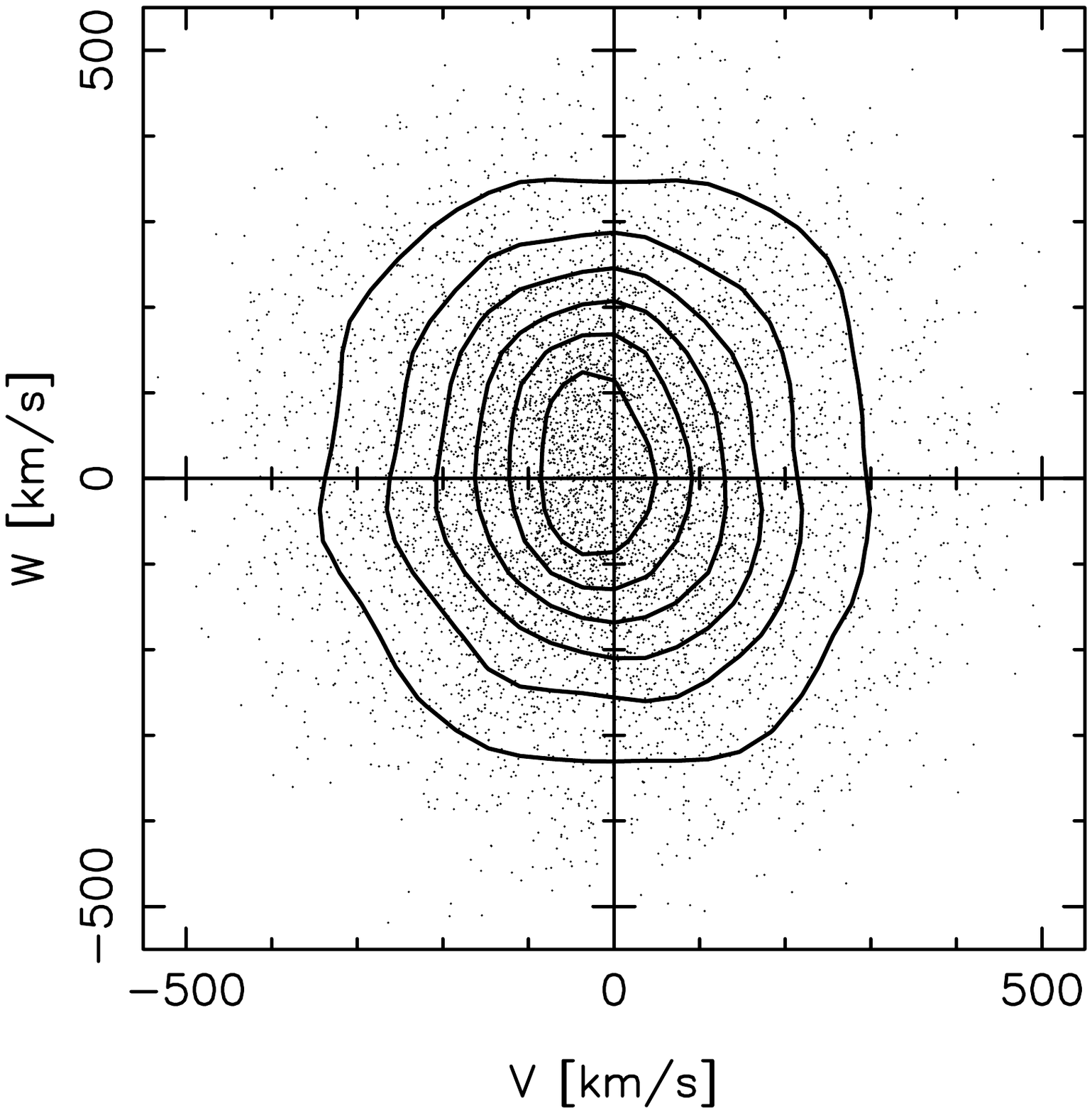}
  \caption{Three projections of the velocity ellipsoids for torus particles in each simulations. The contours correspond to the projected density of particles in each bin while small dots show the actual distribution of particles. 
    }
  \label{fig:velocity_ellipsoids}
\end{figure*}
 
In \reffig{velocity_ellipsoids} we show three projections of the velocity ellipsoid for the dark matter particles in the torus regions of each of the three simulations. We use the Galactic coordinate system with $U$  being the velocity component in the direction of the galactic center, $V$ in the direction of the disk's rotation and $W$ out of the disk plane. Projections of the velocity ellipsoid are shown as three contour plots, each showing  two velocity components ($U, V$ $U,W$ or $V, W$). The contours in each panel show the projected density of dark matter particles. If the velocity distribution in a model were isotropic, the contours in all three panels (in a given column) would be perfectly circular. It is immediately clear that none of the velocity ellipsoids have perfectly circular contours, indicating that all the velocity distributions are anisotropic.

The left hand column shows that the velocity ellipsoid for g1536DM deviates most significantly from spherical in all three projections: in addition the contours (especially the inner most contours) are not very elliptical, are `tilted' and successive contours are strongly misaligned. In addition Model g1536DM  shows the largest variation in contour shape and orientation.  The second column shows that in g1536  the inner most contours in the $U-V$ and $U-W$ planes are significantly tilted at small velocities but the outer contours are more round. The third column shows that for model g15784, all the contours are  flattened along one axis in each projection  but none of the contours are significantly tilted. 

The velocity dispersions of torus particles in \reftab{Simulations} suggest that the ellipsoids for all three simulations deviate significantly from isotropy. The non-circularity of the inner most velocity ellipsoids reflects the fact that at small velocities the distributions deviate strongly from isotropy. On the other hand the near circularity of the outer most contours reflects the fact that at the highest  velocities in each model the velocity distributions become more isotropic. We will see that the difference  in the flattening and tilts of the contours (pointing to velocity anisotropies) affects the speed distribution of dark matter particles and consequently the behavior of the annual modulation signature (see \reffig{ModulationS}). The cause of the tilts in the velocity ellipsoids in g1536DM and g1536 is unclear but could point to an offset between the kinematic symmetry axis of the halo and the assumed plane of the disk (recall that although there is no baryonic disk in g1536DM, it is assumed to lie in the plane perpendicular to the net angular momentum vector of the DM halo.) 

In order to compute the effect of the velocity distribution of DM particles on the direct detection signals we represent the velocity distribution of the DM particles in four different ways, always approximated with an isotropic velocity distribution.
We directly obtain particle velocities from the simulation, and then compare these actual simulation results with three different approximations. In obtaining results directly from the simulation, we do the following: for each DM particle (labeled $k$) in the simulation, we find the velocity and smooth it with a Gaussian kernel\footnote{By turning each delta function in velocity into a narrow Gaussian we more closely reproduce the parent distribution.}. 
For $N_p$ simulation particles of equal mass and velocities $\bv_k$, the sum of Gaussians
\begin{equation}\label{eqn:f}
  \fgal(\bv)
    = \frac{1}{N_p} \sum_{k=1}^{N_p}
        \frac{1}{(2\pi\sigma_s^2)^{3/2}}
        \, e^{-(\bv-\bv_k)^2/2\sigma_s^2}
\end{equation}
can then be taken as a good representation of the velocity distribution of dark matter in the simulated halo.  A softening,
parametrized by $\sigma_s$, is applied to each particle to smooth the
distribution.  We use $\sigma_s = 10/\sqrt{2}$~km/s when showing
velocity distributions.  This choice for $\sigma_s$ is somewhat arbitrary, and was chosen to give a reasonably smooth velocity function without removing too many features.  This value is similar to the order of magnitude of the velocity dispersion observed for dwarf galaxies with masses of the order of the dark matter particle masses used in the simulation (1.11$\times 10^6$\msun).  In addition, for a gravitational softening length of 310 pc used by the simulations, $\sigma_s$ is about the speed below which the actual dark matter particles would be trapped inside the potential well of the dark matter particles of the simulation.  On the other hand we can take $\sigma_s \to 0$ for the mean inverse speed quantity, defined in \refsec{Detection}, that arises in the
calculation of direct detection signals (since $\eta$ is an integrated quantity over velocity, its distribution
is already fairly smooth without further softening required).  
The tilde ($\fgal$) is used to indicate that the distribution is
defined in the galactic (\ie\ non-rotating) rest frame.


For comparison, we then consider three simple empirical approximations to
the velocity distribution, all based upon a truncated Maxwell-Boltzmann (MB)
distribution:
\begin{equation} \label{eqn:TMB}
  \fMB(\bv) = 
    \begin{cases}
      \frac{1}{\Nesc} \left( \pi \vmp^2 \right)^{-3/2}
        \, e^{-\bv^2\!/\vmp^2} , 
        & \textrm{for} \,\, |\bv| < \vesc  \\
      0 , & \textrm{otherwise},
    \end{cases}
\end{equation}
where $z \equiv \vesc/\vmp$, and the quantity
\begin{equation} \label{eqn:Nesc}
  \Nesc = \erf(z) - \frac{2}{\sqrt{\pi}} z e^{-z^2} \, ,
\end{equation}
 is a normalization factor accounting for
the truncation.  The purpose of the truncation is to 
remove particles from the analytic MB distribution (that has tails that extends to infinity), since particles above the escape velocity would have escaped from the galactic potential. This removal is necessary as these high-speed particles are the most likely to produce detectable events in direct detection searches---higher speeds allow for more energy to be deposited in a collision with a nucleus---and inclusion of the tail of the MB distribution would lead to predictions
of signal in some cases where none should be expected (notably, for light WIMPs). This effective treatment of the tail (simply truncating it) is not expected to yield an accurate model of the highest-speed particles; rather, it is intended to avoid the most pathological cases for direct detection analyses. The highest velocity particles are the ones of particular interest near threshold of experiments for low mass WIMPs.  Unfortunately, our simulations simply do not have enough particles in the high velocity tail to allow us to study this regime in detail.

The escape velocity ($\vesc$) is needed to define the three fitted distribution functions.  For each of the different halos, $\vesc$ is set to be the maximum N-body particle speed observed in the corresponding MB's rest frame of the torus.  An alternative definition would be to choose the escape velocity as the highest speed of any particle in the entire halo or the speed of a particle needed to escape the potential from all of the particles in the halo.  We have verified that using one of these speeds instead does not alter any of our conclusions. This is because for our halos, $\vesc$ is large enough (compared to $v_0$) that the observables are largely independent of the exact value used for the escape velocity.  Again, we are not expecting this truncation for the fitted functions to be an accurate representation of the actual N-body dark matter distribution at these high speeds, but instead we are just trying to prevent our fitted velocity distributions (which would extend to arbitrarily large speeds without the truncation) from producing events in a region where there should be none, as dark matter particles with these large speeds would have escaped the galaxy. 

As our first empirical approximation based on this MB form, we consider the 
Standard Halo Model (SHM), the fiducial halo model overwhelmingly assumed in the community
for the purposes of dark matter direct detection phenomenology.  The SHM is an isothermal spherical dark matter distribution \cite{Drukier:1986tm}, which leads to a MB distribution with velocity dispersion $\vmp = \vrot$, where $\vrot$ is the disk rotation speed in the solar neighborhood.  
The disk’s circular velocity in the midplane ($\vrot$) is given by the gradient of the total gravitational potential (dark matter + baryons) at the solar circle. This quantity was computed by Pynbody code \cite{2013ascl.soft05002P} and is given by 
\begin{equation}
\left.\vrot= \sqrt{R\frac{d\Phi}{dR}}\right\vert_{z=0,R=8},
\end{equation}
with $\Phi$ the total gravitational potential (dark matter + baryons).  The Pynbody code averages the above expression around the disk at the specified radius when computing $\vrot$. As $\vrot$ is related to the potential gradient, the SHM velocity distribution is inferred strictly from a simulation's mass distribution and not the actual N-body particle velocities.

We also consider two additional approximations other than the SHM based on fits to the actual N-body particle velocities. We consider MB fits in two cases: assuming a stationary (\ie\ non-rotating) halo (average orbital velocity zero) and allowing for some bulk rotation $\bvbulk$ (average orbital speed $v_{MB}$ obtained from the simulation).   For the stationary halo we simply set the bulk rotation $\bvbulk =0$ and find the best fit MB for $\vmp$ with all of the particle velocities in the simulation.  The result of this fit can be approximated using the most probable speed of the Gaussian distribution 
\begin{equation} \label{eqn:v0Def}
\vmp=\sqrt{\frac{2}{3}}\sigma_v\approx\sqrt{\frac{2}{3}( \sigma_U^2 + \sigma_V^2 + \sigma_W^2)},
\end{equation}
where we have used the standard deviations of the particle velocities to approximate the variance of the MB distribution, $\sigma_v$.  Applying this formula to the velocity standard deviations given in \reftab{Simulations} will approximate the results for $\vmp$ given in \reftab{Models}. For the rotating halo we allow nonzero bulk rotation and find the best fit to both $\bvbulk$ and $v_0$.

We therefore have three
empirical halo models for each simulation: (1) the inferred SHM,
(2) the best-fit MB, and (3) the best-fit rotating MB.  Model parameters
are shown in \reftab{Models}. In \reftab{Models} one can see that the best fit bulk velocity is quite small, at most
$\bvbulk \sim 20$km/sec.  Hence, as mentioned previously in the introduction, we
see no evidence in the simulations of the significant dark matter rotation that would be expected from a ``dark disk''.

\begin{table}
  \begin{center}
  \addtolength{\tabcolsep}{0.5em}
  \begin{tabular}{llccc}
    \hline \hline 
    \hspace*{1.0em}
      & & \textbf{g1536DM} & \textbf{g1536} & \textbf{g15784} \\
    \hline 
    \multicolumn{5}{l}{Standard Halo Model} \\
    & $\vmp (= \vrot)$ [km/s] & 108 & 187 & 273  \\
    & $\vesc$ [km/s]          & 359 & 454 & 600. \\
    \multicolumn{5}{l}{Best-fit Maxwell-Boltzmann} \\
    & $\vmp$ [km/s]           & 139 & 187 & 260. \\
    & $\vesc$ [km/s]          & 359 & 454 & 600. \\
    \multicolumn{5}{l}{Best-fit Maxwell-Boltzmann (rotating)} \\
    & $\vmp$ [km/s]           & 139 & 186 & 259 \\
    & $\vesc$ [km/s]          & 363 & 465 & 602. \\
    & $\bvbulk$ [km/s]        & (0,5.3,0) & (0,21.6,0) & (0,18.5,0) \\
    \hline \hline 
  \end{tabular}
  \end{center}
  \caption[Halo Models]{%
    Empirical halo models used for comparison, each based upon a
    truncated Maxwell-Boltzmann (MB) distribution; see \refeqn{TMB}.
    }
  \label{tab:Models}
\end{table}

\subsection{Results for Velocity and Speed Distributions in Galactic Rest Frame}

In this subsection we discuss the results for the velocity and speed distributions in the rest frame of the galaxy for three N-body simulations: g1536DM  (containing only dark matter particles), g1536 (the same galaxy but includes baryonic physics) and g15784 (a slightly heavier galaxy that also includes baryonic physics).
We illustrate the results for these different distributions produced by the three N-body simulations and also compare them to the three approximations described above that are used to mimic the physics of our Galaxy: the SHM, a MB fit with no rotation and a MB fit with rotation.

\begin{figure*}
  \insertwidefig{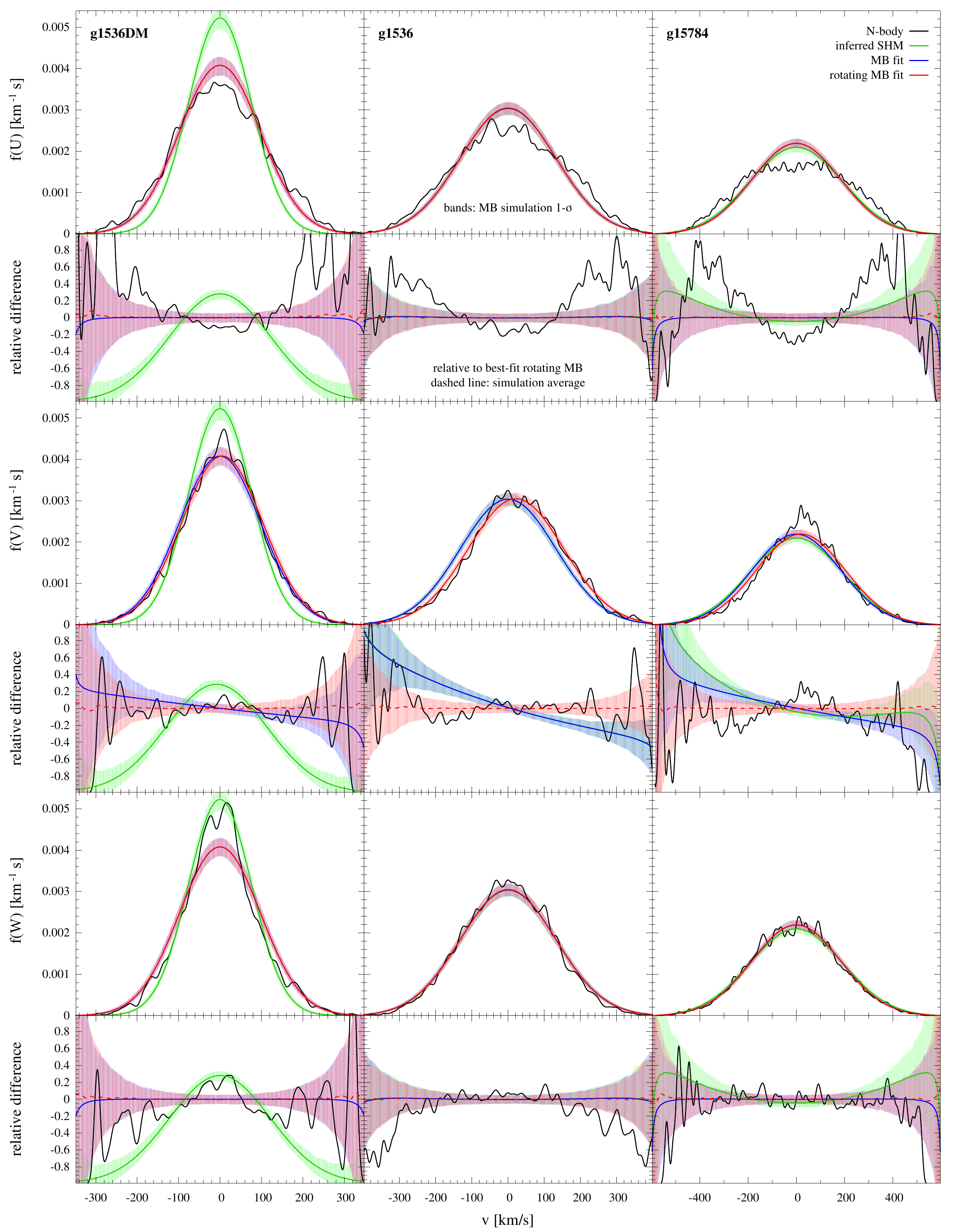}
  \caption{
    \footnotesize
    Velocity component distributions in the galactic rest frame for
    radial velocity $U$ (upper row), azimuthal velocity $V$ (middle
    row), and velocity out of the disk plane $W$ (bottom row).
    Black curves  illustrate results for N-body simulations:  
    g1536DM for dark matter only (left column), g1536 is the same galaxy with
    baryonic physics included (middle column), and g15784 with baryonic
    physics included (right column). 
    Distributions are shown also for inferred SHM (green), best-fit MB
    (blue) and best-fit MB with some bulk rotation (red).  
    Each row has two subpanels:  the actual distribution (upper subpanel);
    and the relative difference to the best-fit rotating MB distribution
    (lower subpanel).
    Colored bands represent the 1-$\sigma$ variation at each $v$ for
    1000 randomly simulated halos with the same number of particles as
    the torus and using the solid lines as the parent populations;
    these are representative of the statistical fluctuations in the
    N-body results due to the limited number of particles.
    }
  \label{fig:GalacticVelDist}
\end{figure*}

The velocity distributions are shown in \reffig{GalacticVelDist} with g1536DM in the left column, g1536 in the middle column and g15784 in the right column.
As in \reffig{velocity_ellipsoids}, we use the Galactic coordinate system to show the components of the velocity distributions: $U$ (upper row), $V$ (middle row), and $W$ (bottom row).  Each of these rows is further subdivided into
two subpanels: the upper subpanel is the actual distribution and the lower subpanel shows 
the relative difference compared to the best-fit rotating MB distribution.  We did not choose the N-body distribution as the standard for relative differences because it fluctuates too much making the plots difficult to interpret.
 Each of the curves for the three fits to the N-body simulations assume a common dispersion in all directions, i.e. spherical symmetry in the velocity ellipsoid.   

All figures display the distributions from the N-body simulation in black, for the inferred SHM in green, the best-fit MB in blue and the best-fit MB with some bulk rotation of the dark matter population in red.  The colored bands represent the 1-$\sigma$ variation at each $v$ for 1000 randomly simulated halos with the same number of particles as the torus and using the solid lines as the parent populations.  These bands are thus representative of the statistical fluctuations in the N-body results due to the limited number of particles.

A general feature to note for all three components of the velocity ($U$, $V$ and $W$) in the case of the DM-only simulation is the fact that
the velocity profile of the SHM is always too narrow (underpredicting the simulated velocity distribution at large negative and positive velocities and overpredicting at the center of the velocity profile). However,
when baryonic physics is included, the agreement between the SHM and the simulated distribution is much better.  We will now examine each of the different cases in more detail.

As observed in the first column, we find the SHM is a poor match to all three velocity components for the N-body simulation that includes only dark matter (g1536DM).  As noted above, the SHM does not match the simulated velocity distribution in any part of the velocity range. The match to the observed distributions improves for the best-fit MB (blue curves) and the best-fit, rotating MB (red curves), although there are still some fairly large discrepancies (note that the blue curves lie entirely below the red curves). 
As the only difference between the SHM and MB models is the value of $v_0$, this means that $v_0$ is underestimated in the SHM for the DM-only simulation.

When baryonic physics is included in g1536 (middle column), we see that all three approximate models are almost indistinguishable and are in good agreement with the simulated velocity distributions.  For this simulation, we find that the best-fit MB is virtually identical to the SHM (i.e. $v_0$ is \textit{not} underestimated in the SHM in the baryonic simulations) and thus the green curve of the SHM is typically hidden behind the blue curve of the best-fit MB.  The distribution of  $V$ is better fitted by a slowly rotating MB model (red curve) than a non-rotating MB model (blue curve). All three models provide good fits for N-body velocity distributions perpendicular to the galactic plane ($W$).

Interestingly, we find that all 3 approximate models do a worse job of fitting the simulated velocity distributions in the plane of g15784 (right column). The model distributions of $U$ are too peaked and the simulated distribution is wider and flatter than any of the models, while the simulated distribution of $V$ is more peaked than any of the models, but $W$ is much better fitted than  the other two velocity components.  The velocity dispersions are $\sigma_U = 205$ km/sec; $\sigma_V=167$ km/sec; and $\sigma_W = 177$ km/sec implying that the velocity ellipsoid is slightly triaxial (as we saw in \reffig{velocity_ellipsoids}).  The assumed isotropy for the approximate models explains the somewhat worse fits in the galactic plane.	


Overall, both the SHM and the MB models provide an acceptable fit to the N-body velocity distributions once baryonic physics is included (i.e. in g1536 and g15784). These fits are much better than when only dark matter is included in the simulation (g1536DM). Since MB distributions are fitted to the N-body velocity distributions it is not surprising that they are a good fit to all three  simulations. However, the SHM only takes as input information about the mass distribution of the dark matter through the circular velocity of the halo  (obtained from the rotation curve).
For an NFW halo \citep{Navarro:1995iw}, the distribution is locally isothermal only at the scale radius $r_s$, where the rotation curve peaks. A typical $\sim 10^{12}~\msun$ N-body halo has $r_s \sim 20$~kpc \citep[e.g.][]{Bullock:2001c}, well outside the Solar circle, so at $R=8$~kpc the rotation curve is still rising, resulting in an underestimate in $v_0$. In contrast, the presence of baryons both increases the steepness of the potential, and causes adiabatic compression of the dark matter, which combine to increase $\vrot$. This results in a rotation curve that peaks at $R \sim 12$~kpc \citep{Klypin:2002aa}, and so $v_0$ is more accurate when measured at the Solar circle. Thus, it is not surprising that the SHM is a poor fit to the dark-matter-only simulation (g1536DM) but is almost indistinguishable from the MB fits in the two simulations with baryons. We will find that these trends will continue for all of the quantities we examine.  

We now comment on a comparison between the results presented for the mass distribution in \reffig{haloshape} and the velocity distributions in \reffig{GalacticVelDist}, for the two galaxies where baryons have
been included in the simulations.  In both galaxies, the addition of baryons has made
the mass distribution more axisymmetric ($b/a$ is driven to 1), and the SHM is a better fit
than without the baryons.  Yet, the case which is the most spherical in terms
of mass distribution (g15784) is the one in which the SHM is a worse fit to the true velocity 
distribution.  We are not certain why this is the case but we may speculate.  The galaxy g15784 is more massive and has experienced a more recent major merger than g1536. Thus, although the mass distribution is more close to spherical, the velocity distribution may not yet have relaxed; it is not yet isotropized. In fact as we saw in \reffig{velocity_ellipsoids} (right) while the velocity ellipsoids are not tilted they are significantly more flattened  at all velocities compared with g1536 which has rounder outer ellipsoids. Since the Milky Way Galaxy is thought to have had a quiet merger history,  the halo of g1536 could be closer in behavior to our own.  To really understand the origin of the differences between these simulations and to draw a direct comparison with the Milky Way, a far larger number of simulations with higher resolution than these is required.  Indeed such simulations are
currently in progress by one of us (Bailin with students) and will be analyzed in the future.  

\begin{figure*}
  \insertwidefig{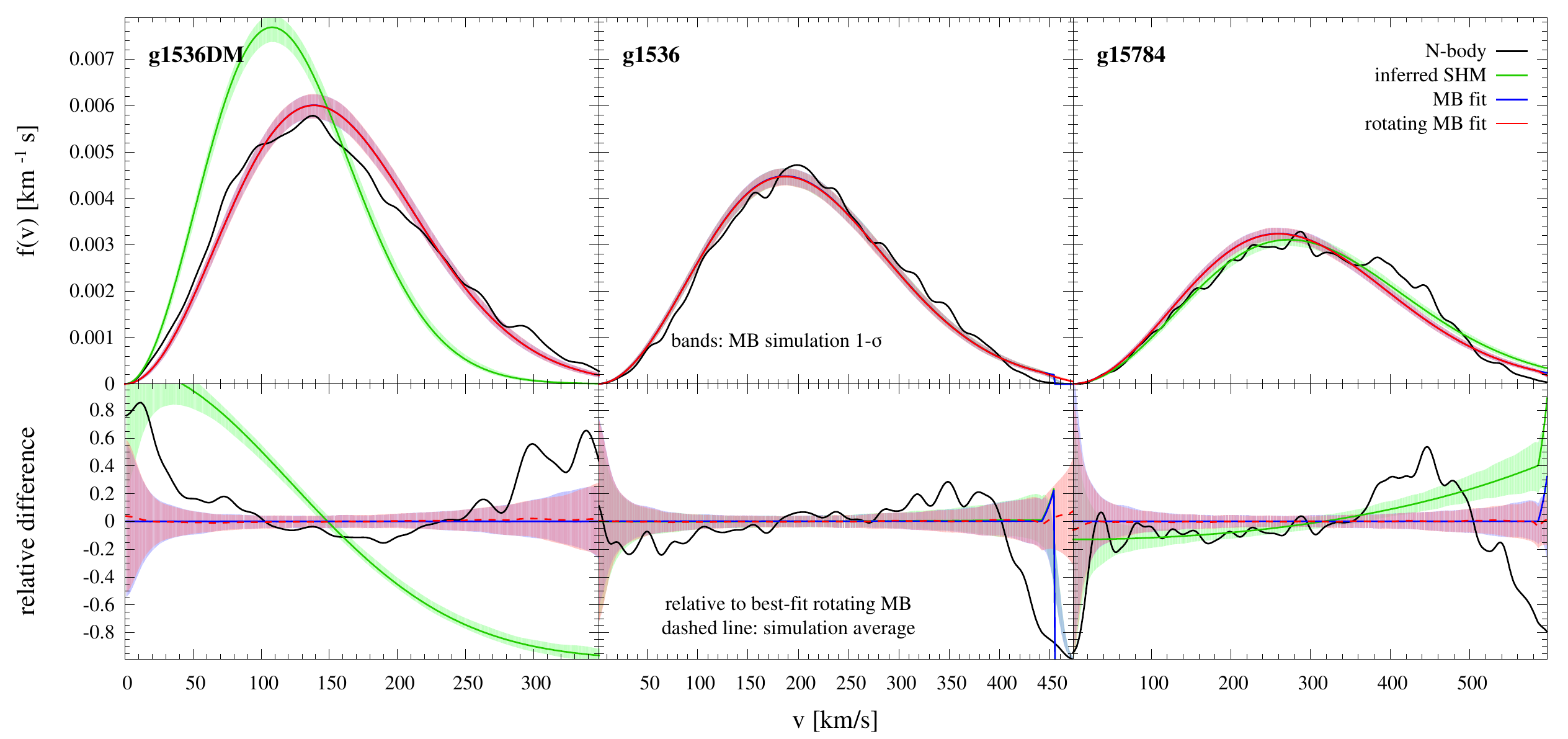}
  \caption{
    (Upper panels) The dark matter speed distribution $f(v)$ in the
    galactic rest frame for the three N-body simulations.
    (Lower panels) Differences in speed distributions (of the N-body simulation and the SHM) relative to the best-fit rotating MB. Curves and bands are the same as described in \reffig{GalacticVelDist}. For g1536 (middle panels) the SHM lies directly below the non-rotating MB (blue). In the two simulations with baryons (middle and right panels)  the deviation of the best-fit rotating MB from the  N-body distribution (black curves) and the SHM (green curves) is less than 25\% except for speeds greater than 400~km/s.}
  \label{fig:GalacticSpeedDist}
\end{figure*}

In \reffig{GalacticSpeedDist} we present the speed distribution in the galactic rest frame\footnote{This distribution depends only
on the magnitude of the velocity, rather than on the individual components of \reffig{GalacticVelDist}.}.  
Again, the three columns refer to the different N-body simulations as above, with the left column for dark matter only while the right two columns have baryons included.  Clearly one observes the SHM is a bad fit to N-body results for the dark matter only simulation (left column).  
Previous work suggesting that SHM is a bad fit to the true DM velocity distribution in our Galaxy is based on similar dark-matter-only simulations \citep{Green:2000jg}.
It is clear from  \reffig{GalacticSpeedDist} however, that including the baryonic physics modifies the velocity distributions in such a way that the SHM is a perfectly adequate description of  the velocity distribution, except perhaps in the high velocity tail where the N-body distribution  has systematically fewer particles with large speeds (this could also simply be due to small number statistics and higher resolution simulations may find better agreement).  


An interesting point to note is that while some rotation is required to fit the distribution of $V$ for g1563, for the speed distributions in the Galactic frame the addition of rotation has negligible impact.  This is evidenced by the fact that the best fit MB and the best fit MB including rotation are almost indistinguishable (the blue curves are behind the red curves).



\section{Direct Detection of Dark Matter}
\label{sec:Detection}

Dark matter direct detection experiments aim to observe the recoil of
a nucleus in a collision with a dark matter
particle~\cite{Drukier:1983gj,Goodman:1984dc,Drukier:1986tm}.
After an elastic collision with a WIMP of mass $\mchi$, a nucleus of
mass $M$ recoils with energy $E = (\mu^2 v^2/M)(1-\cos\theta)$, where
$\mu \equiv \mchi M/(\mchi + M)$ is the reduced mass of the WIMP-nucleus
system, $v$ is the speed of the WIMP relative to the nucleus, and
$\theta$ is the scattering angle in the center of mass frame.  The
differential recoil rate per unit detector mass is
\begin{equation}\label{eqn:dRdE}
  \frac{dR}{dE}(E,t)
    = \frac{\nchi}{M} \, \Big\langle v \frac{d\sigma}{dE}  \Big\rangle
    = \frac{2\rhochi}{\mchi}
      \int d^3v \, v \, \fearth(\bv,t) \, \frac{d\sigma}{dq^2}(q^2,v) \, ,
\end{equation}
where $\nchi = \rhochi/\mchi$ is the local number density of WIMPs and
$\rhochi$ is the local dark matter mass density, $f_{\oplus}(\bv,t)$ is
the time-dependent WIMP velocity distribution in the frame of the detector, and
$\frac{d\sigma}{dq^2}(q^2,v)$ is the velocity-dependent differential
scattering cross-section with the momentum exchange in the scatter
given by $q^2 = 2 M E$.  The time dependence of the velocity
distribution arises from the orbit of the Earth around the Sun and
will be discussed in more detail below.
More detailed reviews of the dark matter scattering
process and direct detection can be found in Refs.~\cite{Primack:1988zm,
Smith:1988kw,Lewin:1995rx,Jungman:1995df,Bertone:2004pz,Freese:2012xd}.

For simplicity, we consider a conventional spin-independent (SI)
coupling of WIMPs to nuclei, for which
\begin{equation}\label{eqn:dsigmadq}
  \frac{d\sigma}{dq^2}(q^2,v)
    = \frac{\sigma_{0}}{4 \mu^2 v^2} F^2(q) \, \Theta(\qmax-q) \, .
\end{equation}
Here, $\Theta$ is the Heaviside step function, $\qmax = 2 \mu v$ is
the maximum momentum transfer in a collision at a relative velocity
$v$, $\sigma_0$ is the scattering cross-section in the
zero-momentum-transfer limit,
and $F^2(q)$ is a form factor to account for the finite size of the
nucleus.\footnote{%
  We use the Helm form factor in our calculations~\cite{Helm:1956zz,Lewin:1995rx}.
  }
For the SI case,
\begin{equation} \label{eqn:sigmaSI}
  \sigmaSI = \frac{4\mu^2}{\pi}
             \Big[ Z \fpSI + (A-Z) \fnSI \Big]^{2} \; ,
\end{equation}
where $A$ is the atomic mass number, $Z$ is the number of protons, and
$\fpSI$ and $\fnSI$ are the effective couplings to protons and
neutrons, respectively.  We furthermore assume identical couplings to
the nucleons ($\fpSI = \fnSI$), which is expected to be approximately
true for many WIMP candidates, in which case the WIMP interaction can
be parametrized in terms of only the scattering cross-section with a
proton $\sigmapSI$, or
\begin{equation} \label{eqn:sigmaSI2}
  \sigmaSI = \frac{\mu^2}{\mup^2}\, A^2 \sigmapSI \; ,
\end{equation}
where $\mup$ is the WIMP-proton reduced mass.  In this scenario, the
dark matter particle is characterized entirely by $\mchi$ and $\sigmapSI$
alone.

For a differential cross-section of the form \refeqn{dsigmadq},
\begin{equation}\label{eqn:dRdE2}
  \frac{dR}{dE}(E,t)
    = \frac{1}{2 \mchi \mu^2} \, \sigma_0 F^2(q) \, \rhochi \eta(\vmin(E),t) \; ,
\end{equation}
where
\begin{equation} \label{eqn:eta}  
  \eta(\vmin,t) = \int_{v > \vmin} d^3v \, \frac{\fearth(\bv,t)}{v}
\end{equation}
is the mean inverse speed, with $\vmin = \sqrt{\frac{ME}{2\mu^2}}$ the
minimum speed of an incoming WIMP for which a nuclear recoil energy $E$
is kinematically allowed.  This restatement of \refeqn{dRdE} allows the
rate to be separated into particle physics terms (the cross-section and
form factor) and astrophysical terms (the local dark matter density and
mean inverse speed).  The impact of realistic halo models on direct
detection results thus requires only an examination of the latter
factors: $\rhochi$ and $\eta(\vmin)$.

In this section, we will examine how the halo models affect the
interpretation of direct detection experimental results in terms of
constraints in the $\sigmapSI-\mchi$ parameter space.  First, though,
we will examine the speed distribution in the Earth's frame and
the resulting mean inverse speed quantity.

\subsection{Speed distribution (Earth frame)}
\label{sec:EarthSpeed}


The velocity distribution in the Earth's frame $\fearth(\bv,t)$ changes
throughout the year due to the motion of the Earth around the Sun.  Assuming $\fgal(\bv)$ is the velocity distribution in the rest frame of the dark matter population, \ie\ the frame where the bulk motion is zero, the velocity distribution in the lab frame is
obtained after a Galilean boost:
\begin{equation} \label{eqn:vdist}
  \fearth(\bv,t) = \fgal(\bvobs(t) + \bv) \, ,
\end{equation}
where
\begin{equation} \label{eqn:vobs}
  \bvobs(t) = \bv_{\odot} + \bV_{\oplus}(t)
\end{equation}
is the motion of the lab frame relative to the rest frame of the dark
matter, $\bv_{\odot}$ is the motion of the Sun relative to that frame,
and $\bV_{\oplus}(t)$ is the velocity of the Earth relative to the
Sun.  For a non-rotating halo, such as
the SHM, $\bv_{\odot} = \bv_{\mathrm{LSR}} +
\bv_{\odot,\mathrm{pec}}$, where $\bv_{\mathrm{LSR}} = (0,\vrot,0)$ is
the motion of the Local Standard of Rest
and $\bv_{\odot,\mathrm{pec}} = (11,12,7)$~km/s is the Sun's peculiar
velocity (see \eg\ Refs.~\cite{Mignard:2000AA, Schoenrich:2009bx} and
references therein).
The $\bV_{\oplus}(t)$ term in \refeqn{vobs} varies throughout the year
as the Earth orbits the Sun, leading to an annual modulation in the
velocity distribution and, thus, the recoil rate.  Written out in
full,
\begin{equation} \label{eqn:voplus}
  \textbf{V}_{\oplus}(t) = 
              V_\oplus \left[
                  \eone \cos{\omega(t-t_1)} + \etwo \sin{\omega(t-t_1)}
                \right] \, ,
\end{equation}
where $\omega = 2\pi$/year, $V_\oplus = 29.8$ km/s is the Earth's
orbital speed around the Sun, and $\eone$ and $\etwo$ are the
directions of the Earth's velocity at times $t_1$ and $t_1+0.25$
years, respectively.  \Refeqn{voplus} neglects the ellipticity of the
Earth's orbit, which is small and only gives negligible changes to the
velocity expression (Refs.~\cite{Green:2003yh,Lewin:1995rx} give the commonly used
more detailed expressions but Ref.~\cite{McCabe:2013kea} provides a corrected form to first order in the eccentricity).  In Galactic coordinates,
\begin{equation}
  \label{eqn:e1e2}
    \eone = (0.9931, 0.1170, -0.01032)
    \quad \text{and} \quad
    \etwo = (-0.0670, 0.4927, -0.8676) \, ,
\end{equation}
where $\eone$ and $\etwo$ are the directions of the Earth's motion at
the Spring equinox (March 21, or $t_1$) and Summer solstice (June 21),
respectively.

\begin{figure*}
  \insertwidefig{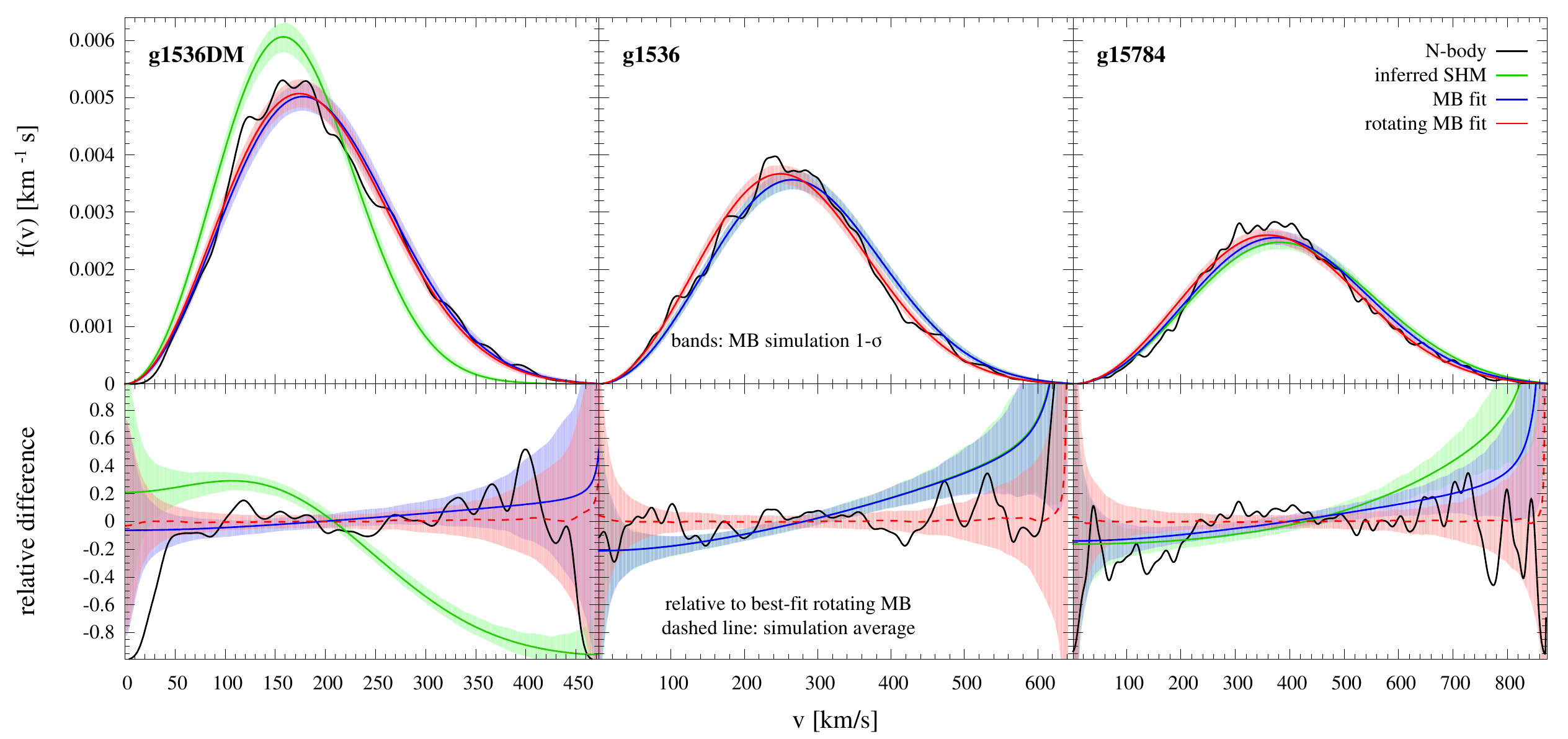}
  \caption{
    (Upper panels) The dark matter speed distribution $f(v)$ as seen
    from Earth.
    (Lower panels) Differences in speed distributions relative to the
    best-fit rotating MB.
    Curves and bands are the same as described in \reffig{GalacticVelDist}.
    }
  \label{fig:EarthSpeedDist}
\end{figure*}


The dark matter speed distribution $f(v)$ as seen from Earth for the three simulations are shown in the upper panels of \reffig{EarthSpeedDist}.  The colors and bands have the same meanings as in \reffig{GalacticVelDist}.  The lower panels of of this figure plot the speed distributions  relative to the best-fit rotating MB.  We again choose the best-fit rotating MB for clarity as the N-body distribution fluctuates too much making the plots difficult to interpret.

As in the case of the Galactic frame, the analysis of the distributions in the Earth's frame  shows that the SHM  is a poor match to the N-body distribution for the case of the dark matter only simulation (g1536DM) 
and the addition of baryonic processes again significantly improves the agreement with the SHM for both g1536 and g15784. 
As previously, the largest deviations are found in the high velocity tail, which might be due to the limitations of the resolution of the simulation which has very few particles with these large speeds.  If these deviations in the high velocity tail are confirmed by higher resolution simulations in the future and are not just a numerical artifact, then the effects on direct detection experiments would be most pronounced if the dark matter particle is light ($\lesssim10$GeV).  This is because particles with that small of a mass will require the highest available speeds to deposit a measurable energy above a typical experiment's threshold.  For higher masses, the experiment would be sensitive to a larger range of the velocity distribution, diluting the effect of the deviations in the tail. 


\subsection{Results for mean inverse speed}
\label{sec:Eta}

\begin{figure*}
  \insertwidefig{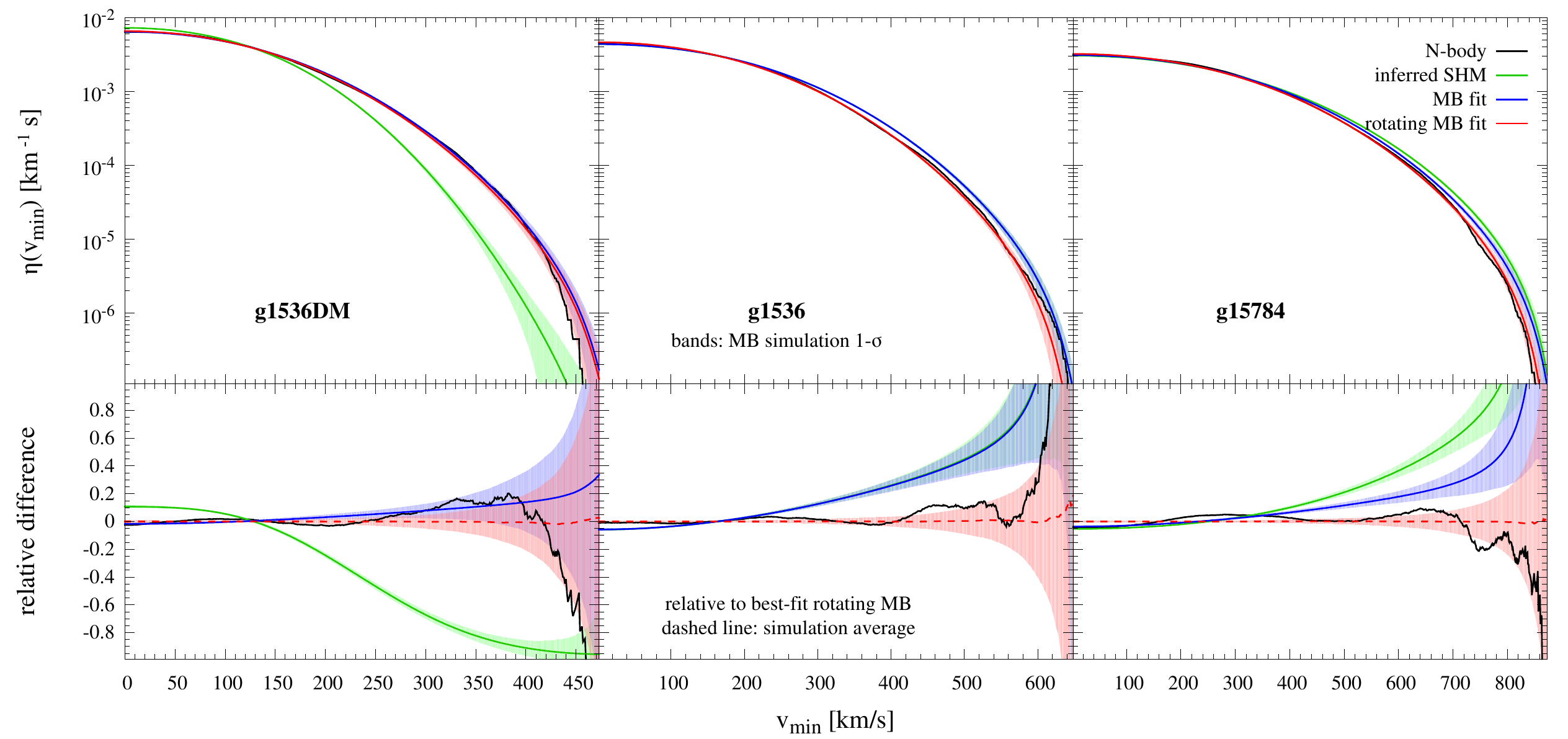}
  \caption{
    (Upper panels) The time-averaged (over one year) mean inverse speed $\eta(\vmin)$
    for the three N-body simulations.
    (Lower panels) Differences in speed distributions relative to the
    best-fit rotating MB.
    Curves and bands are as described in \reffig{GalacticVelDist}.
    }
  \label{fig:Eta}
\end{figure*}

The differential count rate of dark matter scatters in a direct detection experiment is directly proportional to the mean inverse speed $\eta$ of the dark matter particles as given in \refeqn{eta}. In this subsection we study the results from the N-body simulations for this quantity and compare to the same models examined in previous sections.  In \reffig{Eta} we plot the time-averaged mean inverse speed $\eta(\vmin)$ for the three simulations using the same color schemes as in the previous figures.  

As anticipated, the inferred SHM is again a very poor fit to the N-body distribution in the dark matter only simulation (g1536DM).  As expected the best-fit rotating and non-rotating MB do provide very good fits for this case as they are fits to the actual particle velocities.  Once baryonic physics is included in the simulation, the SHM fit improves dramatically but still shows large deviations  (see relative difference plots) at high velocities.  This is true for both g1536 (SHM is again behind the best fit MB band) and g15784.  It is again not clear whether this effect is mostly caused by limitations in the simulation due to small numbers of particles with these higher speeds or an actual inability of the SHM to accurately describe the relevant physics.


\subsection{Results for Amplitude of Annual Modulation}
\label{sec:Sm}

The direct detection signal arising from WIMP dark matter is expected to exhibit an annual modulation due to the motion of the Earth around the Sun. Detection of such an annual modulation was previously described as an important test in establishing the dark matter origin of a signal in direct detection experiments  \cite{Drukier:1986tm}.  In contrast many backgrounds are isotropic and would not exhibit an annual variation.  To describe the annual modulation of the mean inverse speed, we expand it as Fourier series (in time) as defined by
\begin{equation} \label{eqn:Fourier}
  \eta(\vmin,t) = C_0(\vmin) + \sum_{n=1}^{\infty}{C_n(\vmin) \cos{n\omega(t-t_0)}} + \sum_{n=1}^{\infty}{S_n(\vmin) \sin{n\omega(t-t_0)}}.
\end{equation} 

The phase of the expansion ($t_0$) is chosen to be the time of year at which the Earth has the largest relative speed with respect to the dark matter population, in which case $C_1 \propto S_m$, where $S_m$ is the conventional modulation amplitude (see~\refeqn{dRdEmod}).

\begin{figure*}
  \insertwidefig{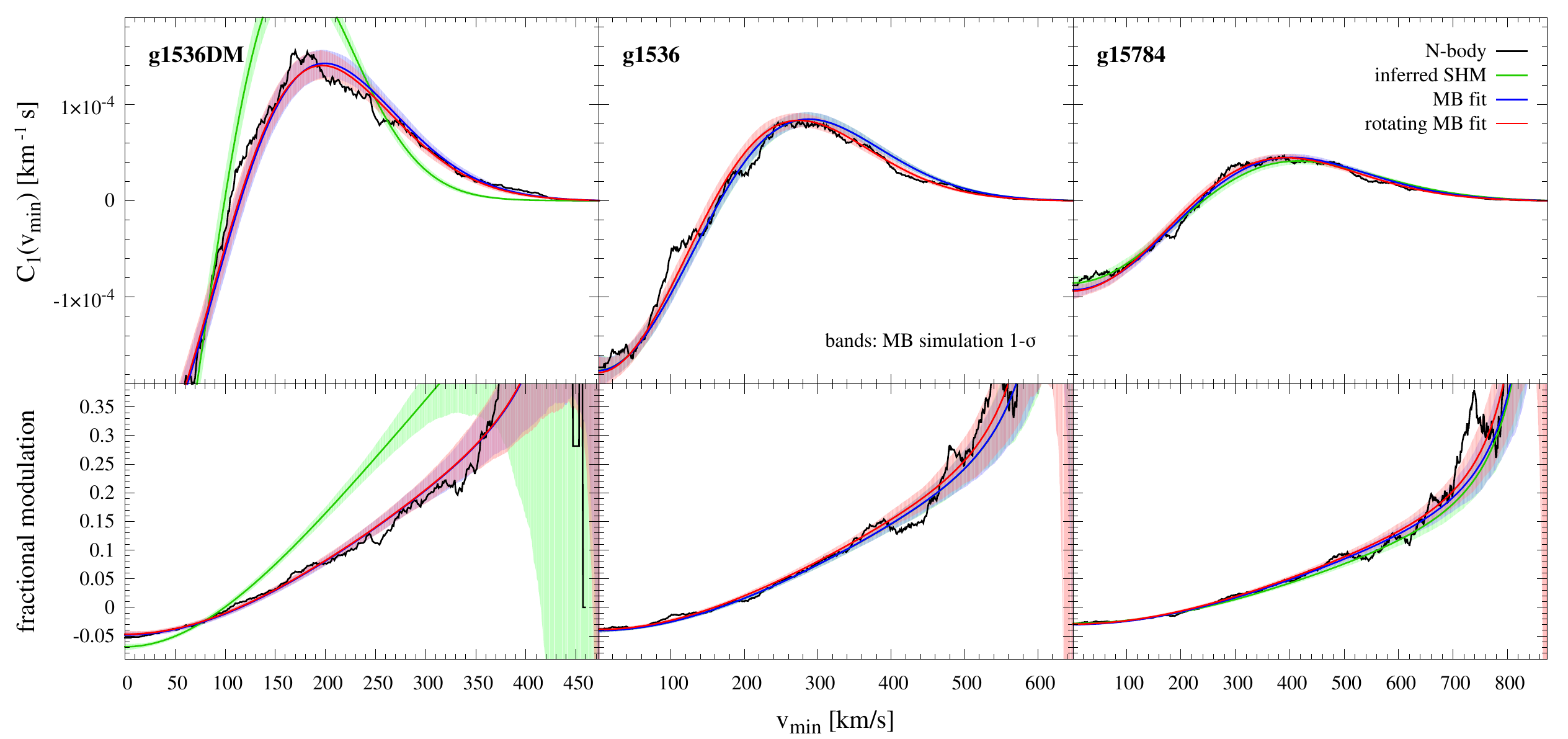}
  \caption{
    (Upper panels)
    The first-order cosine coefficient $C_1$ in the Fourier series
    expansion (in time) of $\eta(\vmin,t)$ for the three N-body simulations.
    The phase of the expansion $t_0$ is chosen to be the time of year
    at which the Earth is moving fastest relative to the dark matter
    population, in which case $C_1 \propto S_m$, where $S_m$ is the
    conventional modulation amplitude.
    Curves and bands are as described in \reffig{GalacticVelDist}.
    (Lower panels)
    The corresponding fractional modulation amplitude,
    \ie\ $C_1(\vmin)/\eta(\vmin)$.
   }
  \label{fig:ModulationC}
\end{figure*}

We now examine how well the SHM reproduces the coefficients in the Fourier series expansion of $\eta(\vmin,t)$ compared to the results of simulations.  We show the fits for $C_1$ in \reffig{ModulationC} with the curves and bands as described in \reffig{GalacticVelDist}.  The bottom panel of these figures illustrates the corresponding fractional modulation amplitude,  \ie\ $C_1(\vmin)/\eta(\vmin)$.  Again as expected, we find the SHM is a very poor fit for the N-body results in the simulation with only dark matter (g1536DM).  However, both the best-fit rotating and non-rotating MB distributions do provide reasonable fits in this case.   All three distributions provide good fits in both g1536 (SHM is again behind the best fit MB band) and g15784, as including baryonic physics in the simulations significantly improves the fit for the SHM.

A dark disk would be defined by a separate population of dark matter confined close to the galactic plane and rotating primarily in the galactic plane.  We found no evidence for a dark disk in our halos, but instead found that a bulk rotation of the entire halo seemed to provide a good description of the velocity distribution.  The amount of rotation found ($\sim$5-20~km/s) is comparable to the 30~km/s of the earth's orbital speed that is responsible for the annual modulation.  One then might expect the modulation signal to be affected, specifically by changing the time of year where the maximum phase occurs.  While this is true for a dark disk \cite{Bruch:2008rx}, a bulk rotation of the halo would have only a negligible effect on the date for the maximum phase.  We indeed find this is the case for our halos with the maximum phase difference created by this rotation being approximately one day.  This is well within the errors of the measurement of the maximum phase for any current or future direct detection experiment. 
    
\begin{figure*}
  \insertwidefig{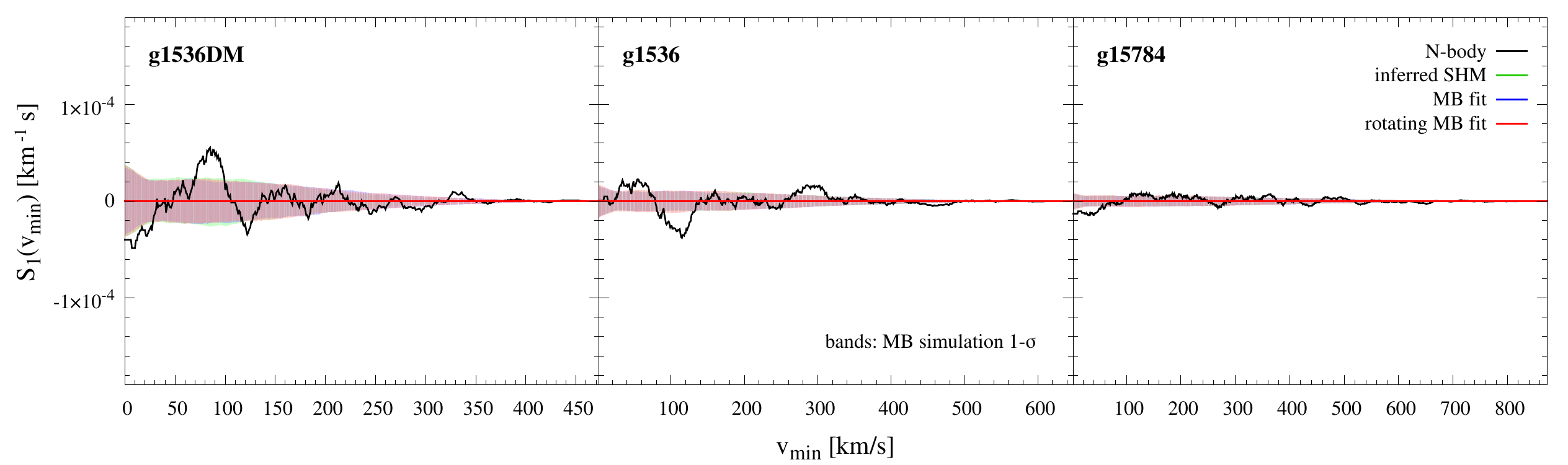}
  \caption{
    The first-order sine coefficient $S_1$ in the Fourier series
    expansion (in time) of $\eta(\vmin,t)$ for the three N-body simulations.
    The phase of the expansion $t_0$ is chosen to be the time of year
    at which the Earth is moving fastest relative to the dark matter
    population, in which case $S_1 \approx 0$ is expected for nearly
    isotropic distributions.
    Curves and bands are as described in \reffig{GalacticVelDist}.
   }
  \label{fig:ModulationS}
\end{figure*}

We also investigated the first-order sine coefficient ($S_1$) for the Fourier expansion of $\eta(\vmin,t)$ for the simulations in \reffig{ModulationS}.  For isotropic dark matter distributions, $S_1 \approx 0$ is expected.  Since the SHM, and both MB fits are by construction isotropic, they all give  $S_1 \approx 0$ as expected with relatively small fluctuations indicated by the bands.  The N-body results (solid black curves), however, show significant deviations from zero, especially for the g1536 simulations both with and without baryons included.  The largest values for $S_1$ are around 30\% of the largest values for $C_1$ because these distributions are significantly non isotropic.  Interestingly, the N-body results for g15784 show very little deviation from $S_1=0$ despite having an anisotropic velocity dispersion. An examination of \reffig{velocity_ellipsoids} suggests that the primary reason for the difference is that the velocity ellipsoids for g15784 are not tilted but those of   g1536DM and g1536 are significantly tilted at small velocities. 

\subsection{Experimental constraints}
\label{sec:Constraints}


The previous sections have focused on the similarities and differences between the velocity distributions for the three simulations we have studied. An equally, if not more, important point to consider is how direct detection experimental results would be affected by the differences among the various halo models. Examining these effects will be the focus of this subsection.

Due to the changing dark matter velocity distribution expected at the Earth as it orbits the Sun, there should be a small ($\sim$ 1-10\%) variation (or modulation) in the recoil rate throughout the year \cite{Drukier:1986tm,Freese:2012xd}. The recoil rate can be described as
\begin{equation} \label{eqn:dRdEmod}
  \frac{dR}{dE}(E,t) = S_0(E) + S_m(E) \cos{\omega(t-t_0)} + \ldots
\end{equation}
where $S_0$ is the average rate, $S_m$ is the modulation amplitude, and higher order terms are often negligible but may become important in certain cases as described in ref.~\cite{Kelso:2013gda}.  Note that $S_m$ can be negative, though this can only occur at low energies.

In the past 25 years many detectors throughout the world have searched for a signal of dark matter interacting in the detector.  The question we want to address is, how reasonable is it to use the SHM or other simple approximations
of the DM velocity distribution in interpreting the data? In particular,
from either positive or null results, we want to extract information about the cross section and mass of a WIMP
that would be provided by such results.  To test the validity of such simplified approximations, we consider three different benchmark detectors.  In these three cases, we examine (a) a positive signal for the average rate $S_0$, (b) a positive signal for the annual modulation $S_m$, and (c) a null result for $S_0$ that would yield an upper limit for the scattering cross section. 

As a benchmark case for a positive $S_0$ signal, we utilize the irreducible spectrum of bulk events presented by the CoGeNT collaboration in Ref.~\cite{Aalseth:2012if}.  The CoGeNT detector, located in the Soudan Underground Laboratory in Northern Minnesota, consists of a 475 grams (fiducial mass of 330 grams) target mass of p-type point contact germanium detector that measures the ionization charge created by nuclear recoils.  We note that this signal, while briefly taken seriously as a signature of WIMPs, is now thought to be due to backgrounds.  Second,
we utilize the DAMA/LIBRA data for the example of a positive signal for $S_m$.  The current DAMA/LIBRA experiment consists of 25 crystals of low-radioactive scintillating thallium-doped sodium iodide (NaI(Tl)) for a total mass of $\sim$250~kg.  The detectors use photomultiplier tubes to collect scintillation light created by nuclear recoils in the crystals.  The total exposure released to date is now 1.33~ton-years and the cumulative signal shows 9.2-$\sigma$ evidence for the presence of an annual modulation~\cite{Bernabei:2013xsa}.  Third, the LUX experiment, located at Sanford Underground Laboratory in South Dakota, is a two-phase (liquid and gas) xenon detector that utilizes photomultipliers to measure both the scintillation and ionization energy created by nuclear recoils.  The LUX collaboration currently has set the strongest scattering cross section upper limit over most of the range of dark matter masses (see Ref.~\cite{Akerib:2013tjd}) and will serve as our benchmark for a null result for $S_0$. 

\begin{figure*}
  \insertwidefig{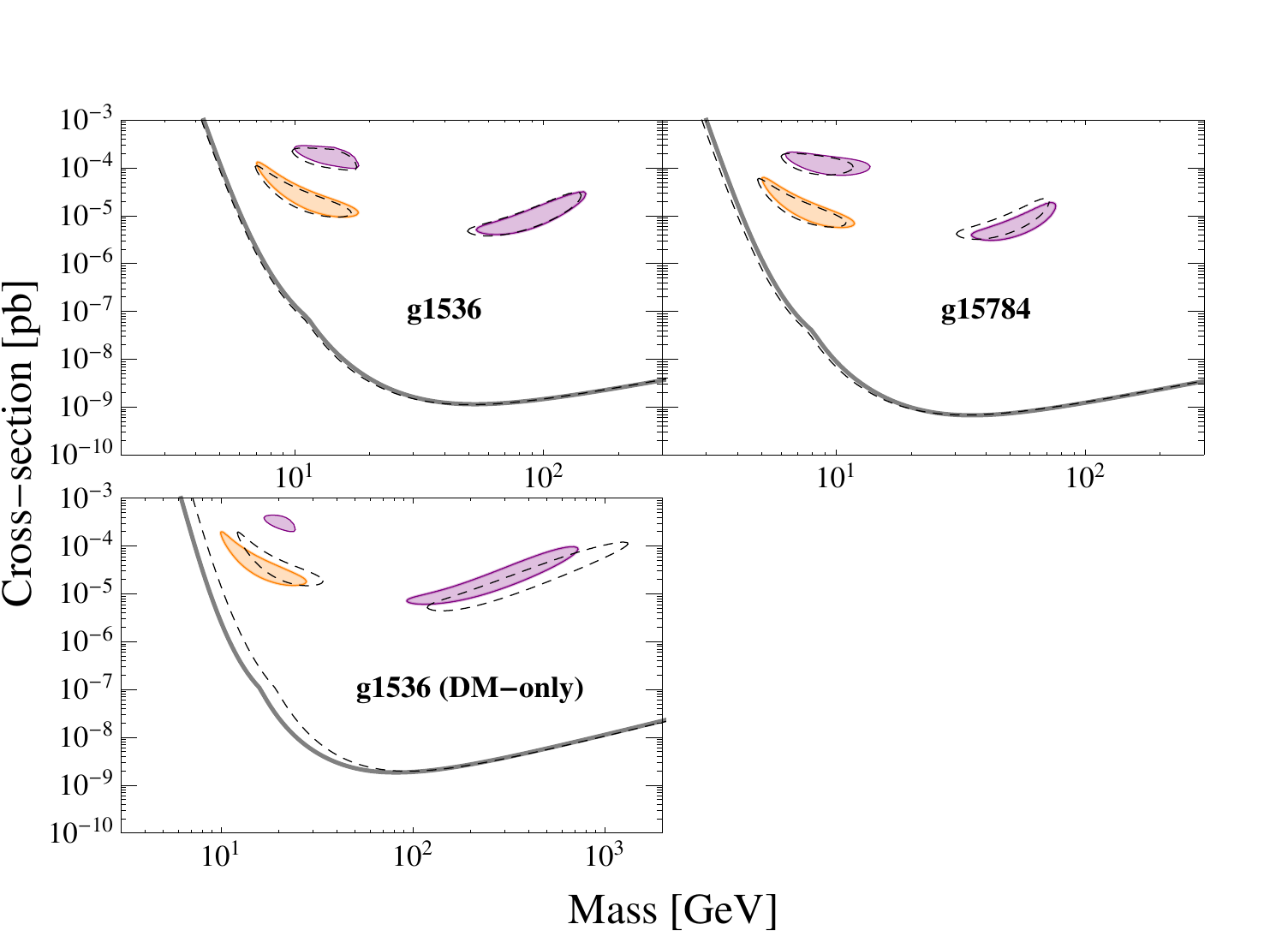}
  \caption{
    The WIMP spin-independent cross-section vs. mass constraints for
    the given N-body simulation using the experimental results from
    CoGeNT (orange region), DAMA (purple region), and LUX (solid gray line).
    CoGeNT and DAMA regions are consistent at the 3-$\sigma$~CL with the
    observed excess events and modulation, respectively, while
    cross-sections above the LUX curve are excluded by at least a
    90\%~CL. We note that the CoGeNT signal is now thought to be due to backgrounds.
    The two separate DAMA regions at $\sim$10~GeV and $\sim$50~GeV are
    due to the relevant scatters being predominantly with sodium and
    iodine, respectively.
    The dashed black curves represent the corresponding regions and limit generated by fitting the data with the inferred SHM for an Earth-like observer in the simulation. In all cases, the local density of dark matter has been fixed to the average dark matter density of the torus as given in \reftab{Simulations}.
  }
  \label{fig:Constraints}
\end{figure*}

We present the results for each of these benchmark cases for the different halo models in the mass, cross-section plane in \reffig{Constraints}.  In each frame of this figure, the colored regions define the 3-$\sigma$ best fit regions for the CoGeNT (orange) and DAMA/LIBRA (purple) data using the $\eta(\vmin)$ produced from the N-body simulation.  The two separate DAMA/LIBRA regions at $\sim$10~GeV and $\sim$50~GeV are due to the relevant scatters being predominantly with sodium and iodine, respectively.  The thick, solid grey line shows the LUX upper limit on the cross section.  The dashed black curves represent the corresponding regions and limit generated by fitting the data with the inferred SHM for an Earth-like observer in the simulation.   Note that the $S_0$ fits (CoGeNT and LUX) use the $\eta(\vmin)$ in \reffig{Eta} while the $S_m$ fit for DAMA/LIBRA uses \reffig{ModulationC}.  The CoGeNT data was collected over a fifteen month period, so the time averaging works well for this data set. For LUX, however, the data was collected over a four month period that included the maximum (about June 1).  Using the time-averaged rate thus leads to a slight under-prediction of the dark matter signal.  This difference produces a negligible change in the limit and can be ignored, however, as many other uncertainties are much more significant.  The results presented in each of the panels also utilize the average dark matter density of the torus in the simulation as given in \reftab{Simulations}.  We chose to show only the inferred SHM for clarity as this model produced the largest difference compared to the N-body simulation results.  

We find the g1536DM simulation (bottom panel) yields the largest difference between the N-body simulation and the reference SHM.  There is a noticeable separation between the N-body and reference SHM regions for both the CoGeNT and high-mass DAMA/LIBRA regions.  Even more striking is the fact that the low-mass region for DAMA/LIBRA has completely disappeared.  In this case, the tail of the modulation spectrum for the inferred SHM decays too quickly to produce a good fit to the DAMA/LIBRA data.  This is not entirely unexpected when considering the somewhat poor fit for $C_1$ by the inferred SHM shown in \reffig{ModulationC}.  This feature supports the commonly held idea that the SHM does not provide a good fit for the dark matter's velocity distribution in galactic halos in {\em dark matter only} N-body simulations.     

As described previously, one of the most interesting results of this study is that the SHM provides a perfectly adequate fit to the simulations with baryons (see \reffig{Eta}~and~\reffig{ModulationC}), defying the ``common lore'' of the field that suggests that the SHM does not adequately describe the dark matter distribution.  This translates to much more similar results for the different regions displayed in \reffig{Constraints} for the g1536 and g15784 simulations.  In particular, there are only very minor differences between the CoGeNT regions for the simulation and the inferred SHM for both g1536 and g15784.  The differences between the simulation and the inferred SHM are more noticeable for the DAMA/LIBRA regions, especially for g15784.  These results are anticipated based on the $C_1$ distributions shown in \reffig{ModulationC}.  This can be understood in terms of the physical nature of the modulation signal.  As long as there is not significant velocity substructure in the vicinity of the Earth, the amplitude of the modulation will be approximately equal to half the difference between the largest yearly scattering rate (time of year when the dark matter-Earth relative velocity is largest) and the smallest yearly scattering rate.  This subtraction amplifies any differences between the two models.  Turning this logic around, the modulation will thus be a sensitive probe of departures from the SHM, i.e. velocity substructures, if they exist within the Milky Way.

Finally, it is also interesting to note that the LUX limits are quite robust to changes in the velocity distribution.  The largest difference again appears in the dark matter only simulation mainly at light masses.  As can be observed in \reffig{Eta}, the high speed tail of the SHM decays much more quickly than the N-body simulation.  For low mass dark matter, only the high velocity tail can produce nuclear recoils above the energy threshold of a detector.  This results 	in fewer expected dark matter scatters, and thus a weaker limit for the SHM at low masses.  At large masses, the two limits are essentially indistinguishable for all of the simulations.  This is expected, as for large dark matter masses, the bulk of the distributions produce scatters above threshold and thus similar numbers of expected events.

\section{Discussion}
\label{sec:Discussion}
A generally accepted tenet among researchers in the dark matter field is that while the SHM provides the general features of the true Milky Way Halo, it does not provide a good quantitative description of the actual distribution of dark matter in our Halo.  A major piece of evidence that supports this claim comes from the poor fit to the distributions in N-body simulations that include {\em dark matter only}.  In our work, we have arrived at the same conclusion for our dark matter only simulation, and find that it is due to an underestimate in $v_0$ due to the still-rising rotation curve near the Sun.  Intriguingly, however, we find that the presence of baryons both increases the steepness of the potential, and causes adiabatic compassion of the dark matter.  These effects combine to bring the peak of the rotation curve near the Sun, and as a result the SHM provides a much better approximation to the simulation data.  In fact, the SHM provides very reasonable fits in both of the simulations that include baryons that we studied.

In addition to investigating the distributions themselves, we also examined the impact these changes would have on the interpretation of direct detection data.  The size of the effects of astrophysics on the interpretation of signals or upper limits for direct detection experiments is something that is not particularly well quantified or even completely understood at this point.  In our work, we have examined how these differences in velocity distributions would translate to different constraints in the mass, cross-section plane for actual experimental data.  In particular, we find that the LUX upper limits are quite robust to these differences even in the dark matter only simulations, especially if the dark matter mass is above $\sim$40~GeV.  For positive signals (either average signal or modulation amplitude), the poor fit of the SHM in the dark matter only simulations leads to somewhat large differences in the mass, cross section plane between the SHM and the N-body simulation.  We again find, however, the simulations including baryonic physics produce only very minor variations in the mass, cross-section plane between the SHM and the N-body simulation.   

We also found no evidence for a dark disk in our halos, but instead found that a small bulk rotation of the entire halo seemed to provide an adequate description of the velocity distribution of the N-body results.  This small bulk rotation of the halo produced only negligible effects on the direct detection observables.  For example, the maximum phase difference for the annual modulation signal created by this rotation was approximately one day.  This is well within the measurement errors of the maximum phase for any current or future direct detection experiment.

In our work, we have used the MaGICC simulations and selected galaxies that are relatively similar to our Milky Way. It would be very interesting to examine other simulations that include different prescriptions for the baryonic physics to determine if these conclusions are generally shared among current simulations.  

On the same day that our work was announced on the arXiv, Borzorgnia and collaborators released a preprint on hydrodynamic simulations of galaxy formation that included baryons for several simulated Milky-Way analogues~\cite{Bozorgnia:2016ogo}.  Their work examined the direct detection implications for these simulated galaxies and they also found that when baryons are included in the simulations, the best fit Maxwell-Boltzmann distribution provides a good fit to the velocity distribution of the dark matter.  Two days later, a preprint by Sloane and collaborators studying four Milky Way-like galaxies with different merger histories was announced on the arXiv~\cite{Sloane:2016kyi}.  This work examined the viability of the SHM, a best fit Maxwell-Boltzmann, and the parametrization of Ref.~\cite{Mao:2012hf}. They found the similar result that including baryons drives the dark matter velocity distribution to be more like the SHM.  This paper also finds that there is a deficit of high velocity dark matter particles compared to the SHM which is different from our results and those of Ref.~\cite{Bozorgnia:2016ogo}.  One possibility for this discrepancy is the strict definition of the SHM used by Sloane et al.  They define the SHM to be a Maxwell-Boltzmann distribution with $\vmp = 220\,$km/s and $\vesc=544\,$km/s, i.e. the parameters derived for the Milky Way.  This is why in Fig.~1 of Ref.~\cite{Sloane:2016kyi}, all of the SHM lines are the same for each of the different simulated galaxies.  One of the fundamental assumptions of the SHM is that the velocity dispersion in the halo is given directly by the rotational speed of the particles, which is a function only of the mass distribution, i.e. $\vmp = \vrot$.  This is why we refer to the model as the ``inferred SHM'' and it is different for each of the halos we have studied (see \reffig{GalacticVelDist} for example).  To determine if this is the root of the discrepancy would require Sloane et al. to define an inferred SHM for their halos using the rotational speed at 8~kpc and compare these inferred SHM to their simulation results.  Independent of this discrepancy, however, it does seem all three works agree that a best fit Maxwell-Boltzmann distribution provides a good fit to the velocity distribution of the dark matter.


\acknowledgments
  C.~S. thanks P.~Sandick and the Department of Physics \& Astronomy at
  the University of Utah for support.
  M.~V. acknowledges support from by University of Michigan's Office of Research,  NSF award AST-0908346, NASA ATP grant NNX15AK79G. M.~V. acknowledges support for HST-AR-13890.001 and J.~B. acknowledges support for program HST-AR-12837, both provided by NASA through a grant from the Space Telescope Science Institute, which is operated by the Association of Universities for Research in Astronomy, Inc., under NASA contracted NAS 5-26555.
   K.~F. acknowledges support from DoE grant DE-SC0007859 at the University of Michigan as well as
   the Swedish Research Council (VR) through the Oskar Klein Centre at Stockholm University.
  CS, KF, and CK also thank the Michigan Center for Theoretical Physics at the
  University of Michigan for support during part of this project.  CK would like to thank CETUP* (Center for Theoretical Underground Physics and Related Areas), for its hospitality and partial support during the 2015 Summer Program. This analysis made use of {\texttt pynbody} package \footnote{https://github.com/pynbody} \cite{2013ascl.soft05002P} which was written by Andrew Pontzen, Rok Ro\v{s}kar, Greg Stinson and others.

\bibliography{ddhalo-final}
\bibliographystyle{JHEP}

\end{document}